\documentclass[twocolumn,pra,floatfix,superscriptaddress]{revtex4-1}
\usepackage{hyperref}
\usepackage{amsmath}
\usepackage{amsfonts}
\usepackage{amsbsy}
\usepackage{xcolor}
\usepackage{soul}
\usepackage{graphicx}

\newcommand{\ket}[1]{\ensuremath{\left\vert #1 \right\rangle}}
\newcommand{\quantmean}[1]{\ensuremath{\left\langle #1 \right\rangle}}

\newcommand{\boldvec}[1]{\ensuremath{\boldsymbol{\mathbf{#1}}}}
\newcommand{\spvec}[1]{\ensuremath{\boldvec{#1}}}
\newcommand{\unitvec}[1]{\ensuremath{\hat{\boldvec{#1}}}}

\newcommand{\pol}{\unitvec{e}}

\newcommand{\Dvhat}{\hat{{\spvec{D}}}}

\newcommand{\Pvhat}{\hat{{\spvec{P}}}}

\newcommand{\dv}{\spvec{d}}

\newcommand{\Dc}{\mathcal{D}}
\newcommand{\Hc}{\mathcal{H}}

\newcommand{\hv}{{\bf h}}

\newcommand{\gv}{{\bf g}}
\newcommand{\eo}{\epsilon_0}

\newcommand{\beq}{\begin{equation}}
\newcommand{\eeq}{\end{equation}}
\newcommand{\bea}{\begin{eqnarray}}
\newcommand{\eea}{\end{eqnarray}}
\newcommand{\<}{\langle}
\renewcommand{\>}{\rangle}
\renewcommand{\(}{\left(}
\renewcommand{\)}{\right)}
\renewcommand{\[}{\left[}
\renewcommand{\]}{\right]}

\newcommand{\commentout}[1]{{}}

\newcommand{\rhoA}{\varrho} 
\newcommand{\rhoB}{\rho_{3D}} 
\newcommand{\rhoC}{\rho} 
\newcommand{\rhoD}{\rho} 
\newcommand{\rhoE}[2]{|\phi_{{#1}}({#2})|^2}  
\newcommand{\rhoF}{\rho} 
\newcommand{\rhoG}{\rho^{\rm tot}} 

\newcommand{\h}{\hat}

\newcommand{\cpsi}{\ensuremath{\hat{\psi}^\dagger}}
\newcommand{\dpsi}{\ensuremath{\hat{\psi}}}
\newcommand{\av}[1]{\ensuremath{\langle #1 \rangle}}
\newcommand{\kappabar}{\ensuremath{\bar{\kappa}}}
\newcommand{\copa}{\ensuremath{\hat{a}^\dagger}}

\newcommand{\dopa}{\ensuremath{\hat{a}}}

\newcommand{\Pop}{\ensuremath{\hat{P}}}
\newcommand{\INT}[1]{\ensuremath{\int \mathrm{d} #1}}
\newcommand{\stochx}{\ensuremath{X}}
\newcommand{\stochxv}{\spvec{\stochx}}

\begin{document}
\author{Mark D. Lee}
\affiliation{Mathematical Sciences, University of Southampton, Southampton SO17 1BJ, United Kingdom}
\author{Stewart D. Jenkins}
\affiliation{Mathematical Sciences, University of Southampton, Southampton SO17 1BJ, United Kingdom}
\author{Yael Bronstein}
\affiliation{Mathematical Sciences, University of Southampton, Southampton SO17 1BJ, United Kingdom}
\affiliation{Universit\'{e} Pierre et Marie Curie, INSP, 4 Place Jussieu, Paris,
F-75005, France}
\affiliation{\'Ecole Normale Sup\'erieure de Cachan, 61 avenue du Pr\'esident Wilson, 94235 Cachan, France}
\author{Janne Ruostekoski}
\affiliation{Mathematical Sciences, University of Southampton, Southampton SO17 1BJ, United Kingdom}
\date{\today }
\title{Stochastic electrodynamics simulations for collective atom response in
optical cavities}
\begin{abstract}
We study the collective optical response of an atomic ensemble confined within a single-mode optical cavity by stochastic electrodynamics
simulations that include the effects of atomic position correlations, internal level structure, and spatial variations in cavity coupling strength and atom density.
In the limit of low light intensity the simulations exactly reproduce the full quantum field-theoretical description for cold stationary atoms and
at higher light intensities we introduce semiclassical approximations to atomic saturation that we compare with the exact solution in
the case of two atoms.
We find that collective subradiant modes of the atoms, with very narrow linewidths, can be coupled to the cavity field by spatial variation of the atomic transition frequency and resolved at low intensities, and show
that they can be specifically driven by tailored transverse pumping beams.  We show that the cavity optical response, in particular both the subradiant mode profile and the
resonance shift of the cavity mode, can be used as
a diagnostic tool for the position correlations of the atoms and hence the
atomic quantum many-body phase.
The quantum effects are found to be most prominent close to the narrow subradiant mode resonances at high light intensities.
Although an optical cavity can generally strongly enhance quantum fluctuations via light confinement, we show that the semiclassical approximation to the stochastic electrodynamics model provides at
least a qualitative agreement with the exact optical response outside the subradiant mode resonances even in the presence of significant saturation of the atoms.
\end{abstract}


\maketitle

\section{Introduction}

The advent of optical cavities revolutionized the study of quantum interactions
of light and atoms by enabling the study of a single quantized light field with
matter~\cite{CarmichaelVol2,Ritsch2013a}.  More recently, advances in trapping cold
atoms~\cite{Kruse2003a,Gupta2007a} in single-mode
high-finesse optical cavities have greatly enhanced the control and tunability of
such systems.  Current experiments are now able to work with quantum degenerate
atoms~\cite{Brennecke2007a,Slama2007a,Colombe2007a,Murch2008a,Brahms2012a,Botter2013a,Zimmermann,Hemmerich} coupling the study of quantum optics with quantum many-body
physics.

Confining many atoms and light inside a cavity can lead to a response that is very different from that of a single atom. The typical signature of such
collective behavior occurs when the emission rates of collective modes are
enhanced (superradiant) or suppressed (subradiant) compared to the emission from
a single atom, and cannot be accounted for by a model of atoms which scatter
light independently.  Within an optical cavity these collective effects can
lead to phenomena such as superradiant lasers~\cite{Bohnet2012a} or realizations
of the Dicke model~\cite{Baumann2010a}.  In multimode cavities, approximate
descriptions of many-atom systems have been shown to result, e.g., in a rich
phenomenology of different quantum
phases~\cite{Gopalakrishnan2009a,Strack2011a}.
The wide variety of physics accessible by cavity-atom systems has necessitated a
number of different
treatments for the studies of many-particle interactions with the cavity field~\cite{Moore1999a,Horak2001a,Gardiner2001a,Nagy2008a,Szirmai2009a,odell2,Mekhov2009a,Niedenzu2013a,Lee2014a,Lee2015a,Sokolov_cavity,odell1,Samoylova:15,Mekhov15,Elliott15,Bhattacherjee15,Larson15,Mazzucchi16,Elliott16},
including studies of cavity optomechanics
\cite{Kippenberg2007a,Kippenberg2008a,Meystre2013a,Aspelmeyer2013a}. Cold atomic ensembles trapped inside or close to the waveguides and fibers form a closely related system where the
light field mediating interactions between atoms is confined in one dimension~\cite{kimblesuper,Rauschenbeutel,kimblefiber,hakuta1,cirackimble,Fan1,Ruostekoski_waveguide}.

Strong collective optical response of cold and dense atomic gases in free space has attracted considerable experimental interest~\cite{Balik2013,Bienaime2010,Pellegrino2014a,wilkowski,Jenkins_thermshift,Jennewein_trans,Ye2016,Guerin_subr16,vdStraten16}. The atoms can exhibit collective radiative resonance
linewidths and line shifts, and experience recurrent scattering where light is scattered more than once by the same atom -- related effects of which have been actively investigated theoretically
in cold atom vapors~\cite{Morice1995a,Ruostekoski1997a,Ruostekoski1997b,Javanainen1999a,Ruostekoski1999a,BerhaneKennedyfer,Clemens2003a,dalibardexp,Jenkins2012a,Castin13,
Javanainen2014a,JavanainenMFT,Kuraptsev14,Bettles_lattice,Lee16,antoine_polaritonic,Longo16,Olmos16,Rey16,Sutherland1D,Jenkins_long,Bettles1D,Sutherland_inverted}.
In free space the light-induced resonant dipole-dipole interactions are of finite range, resulting in light-mediated interactions that depend on the positions of the individual atoms in the ensemble.
The combination of recurrent scattering and the position-dependent light-mediated interactions can lead to a correlated atom response that dramatically differs from that predicted by mean-field theoretical models
of continuous medium electrodynamics in which such effects are ignored~\cite{Javanainen2014a,JavanainenMFT}.

In a cavity the optical response can be quite different from that of free space. Even a single atom can experience recurrent scattering by repeatedly absorbing and emitting the same photon.
Quantum effects can more clearly manifest themselves owing to the directed light confinement.
In contrast to atoms in free space, the sole common electromagnetic mode of a single-mode optical cavity allows long-range scattering of light between atoms without
attenuation.  As a consequence, collective radiative effects can be important in a cavity even if the interatomic spacing is larger than a wavelength.
However, the collective phenomena are simplified due to the absence of attenuation in the scattered light, considerably suppressing the spatial dependence of the light-mediated interactions.
Consequently, in some limits, exact solutions may be obtained in cavities, for example
when all atoms experience an identical cavity field and trapping
potential~\cite{Tavis1968a}. However, modern experiments can involve atomic distributions
which are broad compared to the cavity wavelength. As a result, spatial
variations in coupling strengths, densities and detunings may become
important~\cite{CarmichaelVol2}, leading to position-dependent correlations between individual atoms.

In this paper, we calculate the optical response of a many-atom cavity system where we include the position-dependent correlations between pointlike atoms that are generated by
light-mediated interactions in the cavity. We develop a semiclassical stochastic electrodynamics simulation method that includes recurrent scattering processes between the atoms,
atomic saturation, internal level structure, and the atom statistics. Crucially, the description can incorporate the effects of spatial variation of the
atom distribution, the cavity mode, and the detuning, so that the atoms no longer are assumed to experience identical fields or potentials. We derive the semiclassical approach
from a quantum field-theoretical analysis for the coupled theory of atoms and the cavity field, with an approach analogous to the one we have previously implemented for the interaction of light
with an atomic ensemble in free space~\cite{Lee16}.
In the limit of low light intensity, the stochastic electrodynamics simulation technique  exactly reproduce the full quantum field-theoretical description for cold stationary two-level atoms~\cite{Javanainen1999a,Lee16}.
At higher light intensities where saturation is important, we implement the semiclassical approximation that neglects  quantum fluctuations between the levels.
For quantum degenerate atomic ensembles the quantum statistical position correlations between the atoms can be incorporated in the simulations of the optical response, provided that one can
synthesize the corresponding stochastic ensemble of atomic positions.

Using the stochastic electrodynamics simulations, we discuss the eigenmode structure of the cavity-atom system, with particular emphasis on the collective atomic modes that can be excited.  The most commonly studied optical response of the system can be considered to be due to a collective mode with superradiant characteristics - the cavity optical field induces this collective mode to decay much faster than would be the case for an isolated atom, known as cavity-enhanced spontaneous emission~\cite{CarmichaelVol2}.  We show that, together with this superradiant mode, there exist subradiant collective modes with extremely narrow linewidths, and that these subradiant modes can be coupled to the cavity light by introducing a spatial variation in atomic transition frequency.  The subradiant modes can have distinct spatial profiles, allowing targeted excitation by suitably tailored transverse pump beams, and their long decay times also provide an opportunity for cavity light-storage mediated by collective atomic excitations, provided that the decay to other modes than the cavity mode is suppressed, e.g., by a regular 
arrangement of atoms.

Since our stochastic electrodynamics are semiclassical outside of the low
intensity limit, we also compare such simulations against the exact quantum
treatment obtained by solving the full coupled equations for the atomic
correlation functions governing the optical response~\cite{Ruostekoski1997a} for
a tractable two-atom system. Such comparisons not only inform about the validity
of the numerical approximation, but also reveal quantum features in the two-atom
response of the cavity system.  The semiclassical stochastic simulations are
able to capture the broad behavior of the optical response when it is dominated
by the superradiant collective mode, although we find pronounced quantum
features in the intensity resonance profile in the full quantum treatment, and
these features are beyond the description of the semiclassical treatment.
Quantum effects are even more significant when subradiant modes dominate the
response, and the stochastic simulations do not give a quantitative agreement
outside of the low intensity limit.  For frequencies detuned from collective atomic
mode resonances, the stochastic simulations agree well with the full quantum
results.   The existence of quantum features in the collective cavity-atom
response indicates the tendency of an optical cavity to enhance the effect of
quantum fluctuations via light confinement. 

While the general techniques developed here can be used for any cold-atom system in a cavity, the simulations we present in this paper specialize to systems in which the atoms are additionally confined in a lattice potential along the cavity axis.
A crucial advantage of our simulation technique is that the quantum statistical
position correlations of the atoms are accounted for by means of the stochastic
sampling procedure, a necessary feature to consider for cavities containing
quantum degenerate atomic gases. We show that the optical response can act as a
diagnostic for the atomic many-body phase by considering the different behavior
of atoms in Mott insulator (MI) or superfluid states within the lattice potential.  We show that both the
subradiant spectrum and the resonance shift of the cavity mode are appreciably
altered by the different position correlations of those states. By comparing 
the coherently and incoherently scattered light we also find how the difference between the total
scattered light intensity and the coherently scattered light intensity is sensitive to the atom statistics,
indicating how the fluctuations of the atomic positions are mapped onto the fluctuations of the scattered light.

We begin in Sec.~\ref{sec:cavityformalism} by deriving the full quantum field-theoretical formalism for
atom-cavity system.  Specializing in
Sec.~\ref{sec:cavitytwolevel} to a system of two-level atoms, we discuss the
stochastic method that allows an efficient numerical simulation.  In Sec.~\ref{sec:eigenmodes} we calculate the eigenmode structure of
the cavity-atom system, before studying the two atom case and the comparison
between stochastic simulations and the full quantum field-theoretical treatment
in Sec.~\ref{sec:numericaltwoatomscavity}.  Finally, in Sec.~\ref{sec:manybody}
we show how the different atomic many-body correlations in MI and
superfluid states can easily be accounted for in the stochastic simulations, and
their effects on the optical response of the system.

\section{Hamiltonian and equations of motion}
\label{sec:cavityformalism}

\subsection{System}

We consider atoms confined within an optical cavity of length $L$. We
assume tight confinement in the transverse dimensions such that the motion of the
atoms is restricted to a single transverse mode -- one which is sufficiently
narrow that the coupling of atoms to the cavity depends only upon the
longitudinal position along the cavity axis $x \in (0,L)$. The system then
reduces to a one-dimensional (1D) problem for atomic quantities integrated over the
transverse dimensions, such as the atomic polarization $\Pvhat(x)$. We include
two pumping mechanisms, both at frequency $\Omega$, a direct cavity axis pump of
strength $\eta$ and a beam illuminating the atoms in the cavity from a transverse
direction with a strength, shape and polarization incorporated into $\hv(x)$.

Within the dipole approximation, and expressing the Hamiltonian in the
\emph{length} gauge from the Power-Zienau-Woolley
transformation~\cite{PowerZienauPTRS1959,Woolley1971a,CohenT}, the interaction
between the atoms and cavity light modes may be given by the 1D Hamiltonian
density
\beq
\Hc_{int}=-{1\over\eo}\,\Pvhat(x)\cdot\Dvhat(x) \,.
\label{eq:Hint}
\eeq
Here, $\Dvhat(x)$ is the electric displacement, which is the basic dynamical
variable for light, and which can be separated into a sum of the positive and
negative frequency components $\Dvhat^+(x)$ and $\Dvhat^-(x)$ respectively, with
$\Dvhat^+(x) = [\Dvhat^-(x)]^\dagger$.  The positive frequency component can be
written in terms of the cavity mode photon annihilation operators $\dopa_q$
\beq
\Dvhat^+(x) = \sum_q\zeta_q\pol_q\dopa_q(t)f_q(x)\, ,\quad \zeta_q=\sqrt{{\hbar
\epsilon_0\omega_q\over 2 L}}\,,
\eeq
where $\pol_q$ is the mode polarization, and $f_q(x)$ encapsulates the spatial
variation of the cavity mode.  We specialize now, and in the remainder of this
paper, to the case of a cavity with a single mode relevant mode, and for clarity
drop the index $q$, however the treatment that we subsequently develop could, in
principle, be extended to treat multimode cavities.

Defining  $\gv(x) = \zeta f(x)\pol/\hbar\eo$, such that for a Fabry-Perot
cavity $\gv(x)=\gv_0 \sin(qx)$, while for a ring cavity we may have modes $\gv(x) = \gv_0
e^{\pm iqx}$, then the Hamiltonian density of Eq.~\eqref{eq:Hint} becomes
\beq
\Hc_{int}\approx-\hbar\left[\gv(x)\cdot\Pvhat^-(x)\dopa(t)
+\gv^*(x)\cdot\Pvhat^+(x)\copa(t)\right]\,.
\label{eq:JCHamiltonian}
\eeq
Here we have additionally made the rotating wave approximation.  We note that, in
principle, counter-rotating terms could be included in a treatment analogous to
that below, and may be required to consider very strong coupling.

Factoring out the dominant driving frequency time dependence for all
time-dependent operators, the full Hamiltonian for the system in the rotating
frame of the pump may be written~\footnote{In length gauge of the
Power-Zienau-Woolley transformation, there is in addition a term proportional to
$\Pvhat(x)\cdot \Pvhat(x)$, however we neglect this term here since for
non-overlapping point dipoles this it vanishes.}
\begin{align}
H &=\INT{x}\left(\Hc_g+\Hc_e\right) \nonumber \\
&-\hbar\int \mathrm{d}x
\,\hv(x)\cdot\Pvhat^-(x)+\hv^*(x)\cdot\Pvhat^+(x) \nonumber \\
&-\hbar\int \mathrm{d}x
\left[\gv(x)\cdot\Pvhat^-(x)\dopa+\gv^*(x)\cdot\Pvhat^+(x)\copa\right] \nonumber \\
&-\hbar\Delta_c\copa\dopa-\hbar\eta\left(\dopa+
\copa \right)\,,
\end{align}
where $\Delta_c = \Omega-\omega_c$ is the detuning of the cavity mode from the
pump.  Here the second line includes the transverse pumping of the atoms, and the
last line describes the cavity mode and its direct pumping with strength $\eta$.
In the first line $\Hc_g$ and $\Hc_e$ are Hamiltonian densities for atomic fields
governing the ground and excited levels respectively.  In many realistic
experimental situations the internal sublevel structure may allow multiple
transitions to be driven by the cavity mode.  We include the sublevel structure
by defining quantum field operators $\dpsi_{g\nu}(x)$ and $\dpsi_{e\eta}(x)$ for
the sublevels involved in the transition $\ket{g,\nu}\rightarrow\ket{e,\eta}$.
For the case of the linear Zeeman shift caused by a magnetic field of strength
$B(x)$ the effective 1D atomic Hamiltonian densities are then given by
\begin{align}
\Hc_g & =\hat\psi^\dagger_{g\nu}(x)\big( {\mu_BB(x)}  g_l^{(g)}\nu \big)
\hat\psi_{g\nu}(x)\,,\label{eq:Hg}\\
\Hc_e & =\hat\psi^\dagger_{e\eta}(x)\big( {\mu_BB(x)}  g_l^{(e)}\eta
+\hbar\omega_0 \big) \hat\psi_{e\eta}(x)\,,\label{eq:He}
\end{align}
where $g^{(g)}_l$ and $g^{(e)}_l$ are the Land\'{e}
g-factors~\cite{BransdenJoachain} for levels $g$ and $e$, and $\omega_0$ is the
resonance frequency of the $\ket{g}\leftrightarrow\ket{e}$ transition in the
absence of any magnetic field.

In this notation, the polarization operator $\Pvhat^+$ can be represented as a
sum over contributions from the different possible transitions
\begin{align}
\Pvhat^+(x) &= \sum_{\nu,\eta} \dv_{g\nu e\eta}\hat\psi^\dagger_{g\nu}(x)
\hat\psi_{ e\eta}(x) \equiv \sum_{\nu,\eta}
\Pvhat^+_{\nu\eta}(x)\,,
\label{pol}\\
\Pvhat^+_{\nu\eta}(x) &\equiv \dv_{g\nu e\eta}\hat\psi^\dagger_{g\nu}(x)
\hat\psi_{ e\eta}(x)\,,
\label{polcomp}
\end{align}
where $\dv_{g\nu e\eta}$ represents the dipole matrix element
for the transition $|e,\eta\>\rightarrow |g,\nu\>$
\beq
\dv_{g\nu e\eta}\equiv \Dc \sum_\sigma \pol_{\sigma}
\< e\eta;1g|1\sigma;g\nu\> \equiv \Dc \sum_\sigma \pol_{\sigma} {\cal C}_{\nu,\eta}^{(\sigma)} \,.
\label{dipole}
\eeq
Here the sum is over the unit circular polarization vectors $\sigma=\pm1,0$, and
${\cal C}_{\nu,\eta}^{(\sigma)}$ denote the Clebsch-Gordan coefficients of the
corresponding optical transitions. The reduced dipole matrix element is
represented by ${\cal D}$
(here chosen to be real) and $\dv_{e\eta g\nu}=\dv_{g\nu e\eta}^*$. The light
fields with the polarizations $\pol_\pm$ and $\pol_0$ drive the transitions
$|g,\nu\>\rightarrow|e,\nu\pm1\>$ and $|g,\nu\> \rightarrow|e,\nu\>$,
respectively, in such a way that only the terms $\sigma=\eta-\nu$ in Eq.~\eqref{dipole}
are nonvanishing.

The open nature of the cavity system is important to consider, and the effect of
transmissive loss of photons from the cavity mode through the end mirrors at a
rate $\kappa$ may be included by considering the evolution of the density matrix
$\h \rhoA$
\beq
{d \h \rhoA\over dt} = {1\over i\hbar} \[ H,\h \rhoA\] + \kappa \( 2 \h a \h \rhoA \h a^\dagger - \h a^\dagger \h a \h \rhoA -\h \rhoA \h a^\dagger \h a \)\,.
\eeq
The equation of motion for $\h \rhoA$ can be used to derive equations of motion
for the cavity field $\dopa(t)$ also incorporating, e.g., the backaction of
quantum measurement~\cite{CarmichaelVol2}.

\subsection{Equations of motion for quantum field operators}

\subsubsection{Light}

We may now calculate equations of motion for the atomic and cavity mode
operators.  From the equation of motion for $\dopa$, and
assuming that $\kappa$ is large compared with the other relevant frequency
scales of the system, we adiabatically eliminate the cavity field. In the limit
that fluctuations from the emitting atoms dominate, we obtain the dependence of
$\dopa$ on the atomic polarization
\beq
\dopa = \dopa_F + \frac{i}{\kappabar}\int \mathrm{d}x \Pvhat^+(x)\cdot\gv^*(x)
\label{eq:cavityeqnfora},
\eeq
where
\beq
\frac{1}{\kappabar} = \frac{\kappa+i\Delta_{c}}{\kappa^2+\Delta_{c}^2},
\eeq
and $\hat{a}_F$ represents the free field that would exist if there were no
atoms present.

\subsubsection{Matter}

Substituting Eq.~\eqref{eq:cavityeqnfora} into the equations of motion for the
atomic operators, and assuming that the atomic motional state is unchanged by
the scattering of light~\footnote{we therefore neglect any change to the atomic
motional energy from a scattering event, an assumption equivalent to assuming
stationary atoms throughout.  See Ref.~\cite{Lee16} for a more detailed
discussion of this approximation}, leads to
\begin{align}
\dot{\dpsi}_{g\nu}(x,t) &=
i\Delta_{g\nu}\dpsi_{g\nu} +i\hv^*(x)\cdot\dv_{g\nu e\eta}
\dpsi_{e\eta}(x)
\nonumber \\
&+\frac{1}{\kappabar^*}\INT{x'} \dv_{g\nu e\eta}\cdot{\sf G}^*_c(x,x')
\Pvhat^-(x')\dpsi_{e\eta}(x) \nonumber \\
&+i\gv^*(x)\cdot\dv_{g\nu e\eta}\hat{a}_F^\dagger\dpsi_{e\eta}(x),\label{eq:cavityeqnmotionpsig} \\
\dot{\dpsi}_{e\eta}(x,t) &=
i\Delta_{e\eta}\dpsi_{e\eta}
+i\hv(x)\cdot\dv_{e\eta g\nu} \dpsi_{g\nu}(x)
\nonumber \\
&-\frac{1}{\kappabar}\INT{x'}\dv_{e\eta g\nu}\cdot{\sf G}_c(x,x')\Pvhat^+(x')
\dpsi_{g\nu}(x) \nonumber \\
&+i\gv(x)\cdot\dv_{e\eta g\nu}\dpsi_{g\nu}(x)\hat{a}_F \nonumber \\
&-\frac{1}{\kappabar}\dv_{e\eta g\tau}\cdot{\sf G}_c(x,x)
\dv_{g\tau e\zeta}\dpsi_{e\zeta}, \label{eq:cavityeqnmotionpsie}
\end{align}
where repeated indices $\zeta,\tau$ are summed over. The atom-light detuning is
denoted by the expressions
\begin{align}
\Delta_{e\eta}=&\Omega-(\omega_0+{\mu_BB}  g_l^{(e)}\eta/\hbar) \, ,\\
\Delta_{g\nu}=&-{\mu_BB}  g_l^{(g)}\nu/\hbar\, ,
\end{align}
where we have again factored out the dominant pump frequency $\Omega$.  We
have also defined the cavity propagator
\beq
{\sf G}_{c}(x,x') = \gv(x)\gv^*(x'). \label{eq:cavitypropagator}
\eeq
Due to the
ability of the cavity to mediate long-range photon exchange between atoms this
propagator depends only upon the positions $x$ and $x'$, rather than the distance
$|x-x'|$ which would appear in free-space~\cite{Lee16}.
Additionally, we have reordered the free-field contribution $\hat{a}_F$
($\hat{a}_F^\dagger$) to the right (left) hand side. We will assume the free
field to be in a coherent state and this ordering means that the free-field
operators will appear as multiplicative classical coherent fields
$\av{\hat{a}_F} = a_F =  i\eta/\kappabar$ after expectation values are taken.
The commutator \beq \left[\dpsi_{g\nu}(x),\dopa_F\right] = -\frac{i}{\kappabar}
\gv^*(x)\cdot\dv_{g\nu e\eta}\dpsi_{e\eta}(x)\,, \label{eq:cavitycommutator}
\eeq which is required for the reordering is readily obtained from
Eq.~\eqref{eq:cavityeqnfora} by observing that atomic operators must commute
with $\dopa$ at equal times.

\subsection{Cavity-atom optical response for two-level atoms}

\subsubsection{The general two-level atom case}

Finally, combining the above equations of motion for atom fields, and reordering the
free-field contributions, results in the coupled equations for the
polarizations and one-body correlation functions which together give the
optical response of the system.  We give the full expressions in
App.~\ref{App:AppFullEqns}, but for clarity we now simplify to the case of
atoms which are well described by a two-level model, coupling to a single mode
optical cavity, a case which we use in the remainder of the paper unless
otherwise specified.  The two-level optical response is governed by the coupled
equations
\begin{align}
{d\over dt}\,\Pop^+(x) &=
\left(i\bar{\Delta}- \frac{{\cal D}^2}{\kappabar}G_c(x,x)\right)\Pop^+(x) \nonumber \\
&+
\frac{{\cal D}^2}{\kappabar}\INT{x'} G_c(x,x')\,\cpsi_e(x)\Pop^+(x')\dpsi_e(x)
 \nonumber \\
&- \frac{{\cal D}^2}{\kappabar}\INT{x'} G_c(x,x')\,\cpsi_g(x)\Pop^+(x')\dpsi_g(x)
\nonumber \\
&-i{\cal D}^2\left[\cpsi_e\dpsi_e-\cpsi_g\dpsi_g\right] \left[ g(x)\hat{a}_F +h(x)\right]
\,,\label{eq:cavitydPdt}\\
{d\over dt}\,\cpsi_{e}\dpsi_{e} &= -{d\over dt}\,\cpsi_{g}\dpsi_{g}\,, \\
&=-2{\cal D}^2\mbox{Re}\left[\frac{1}{\kappabar}\right]G_c(x,x)\cpsi_e\dpsi_e \nonumber \\
&-\frac{{\cal D}}{\kappabar}\INT{x'}G_c(x,x')\,\cpsi_e(x)\Pop^+(x')\dpsi_g(x)\nonumber \\
&-\frac{{\cal D}}{\kappabar^*}\INT{x'}G^*_c(x,x')\,\cpsi_g(x)\Pop^-(x')\dpsi_e(x)\nonumber \\
&+ig(x)\Pop^-\hat{a}_F -ig^*(x)\hat{a}_F^\dagger\Pop^+ \nonumber \\
&+i\left[h(x)\Pop^--h^*(x)\Pop^+\right]\,, \label{eq:cavitydrhogdt}
\end{align}
where we have defined the two-level atom versions of Eqs.~\eqref{pol} and
\eqref{eq:cavitypropagator} by $\Pop^+ = {\cal D}\cpsi_g\dpsi_e$ and $G_c(x,x') =
g(x)g^*(x')$, with $g(x) = \gv(x)\cdot\dv_{eg}/{\cal D}$ and $h(x)
=\hv(x)\cdot\dv_{eg}/{\cal D}$.  The detuning $\bar{\Delta} = \Delta_e-\Delta_g$
now represents simply the detuning between the pump frequency and the atomic
resonance frequency.  In the presence of a spatially dependent applied magnetic
field strength, $\bar{\Delta}$ can be spatially dependent [see Eqs.~\eqref{eq:Hg}
and~\eqref{eq:He}], however for brevity we will not generally indicate the
spatial dependence unless relevant.

In this two-level model, the origin of the terms in
these equations may be discussed transparently.  In Eq.~\eqref{eq:cavitydPdt},
the last line represents the interaction of an atom at point $x$ with the
transverse driving field $h(x)$ and the driven field of the cavity $\dopa_F$.
The effect of saturation is accounted for by the factor of
$(\cpsi_e\dpsi_e-\cpsi_g\dpsi_g)$.  The second and third lines include the
effect of a dipole at point $x'$ interacting with an atom at point $x$ via the
cavity mode, an interaction governed by the propagator $G_c(x,x')$.  Finally,
the term proportional to $G_c(x,x)$ in the first line arises from the reordering
of the free-field contribution.  In free-space, the similar procedure resulted in
terms corresponding to both the spontaneous emission and the Lamb
shift~\cite{Lee16}.  In the cavity, the term has the form of a
self-interaction of a dipole at $x$ via the cavity field.  However, the real and
imaginary parts could alternatively be expressed as the position dependent
cavity-enhanced emission rate ${\cal D}^2|g(x)|^2\kappa/(\kappa^2+\Delta_c^2)$
and Lamb shift ${\cal
D}^2|g(x)|^2\Delta_c/(\kappa^2+\Delta_c^2)$~\cite{Brune1994a}, respectively.

\subsubsection{The limit of low light intensity}

The system simplifies greatly in the limit of low light intensity, where we may
work to first order in the excited level operator $\dpsi_{e\nu}$ or driving
field operator $\dopa_F$.  In this limit, the excited state density operator
vanishes, the ground state density operator is invariant, and it remains only
to treat the equation for the atomic
polarization operator, which becomes
\begin{align}
{d\over dt}\,\Pop^+(x) &=
\left(i\bar{\Delta}- \frac{{\cal D}^2}{\kappabar}G_c(x,x)\right)\Pop^+(x) \nonumber \\
&- \frac{{\cal D}^2}{\kappabar}\INT{x'} G_c(x,x')\,\cpsi_g(x)\Pop^+(x')\dpsi_g(x)
\nonumber \\
&+i{\cal D}^2\cpsi_g\dpsi_g \left[ g(x)\hat{a}_F +h(x)\right]
\,.\label{eq:cavitydPdtlowint}\\
\end{align}
Taking expectation values of both sides of this equation gives the equation of
motion for the polarization $P(x) \equiv \av{\hat{P}^+(x)}$, from which the optical
response is readily calculated from Eq.~\eqref{eq:cavityeqnfora}.  However, the
equation of motion for $\av{\Pop^+(x)}$ depends on the two-body correlation
\beq
P_2(x;x') \equiv \av{\cpsi_g(x)\Pop^+(x')\dpsi_g(x)}\, .
\eeq
The equation of motion
for $P_2$ may be derived in an analogous manner
\begin{align}
{d\over dt}&\,P_2(x;x')=
\left(i\bar{\Delta}- \frac{{\cal D}^2}{\kappabar}G_c(x',x')\right)P_2(x;x')\nonumber \\
&-\frac{{\cal D}^2}{\kappabar}G_c(x',x)P_2(x';x) \nonumber \\
&- \frac{{\cal D}^2}{\kappabar}\INT{x''}
G_c(x',x'')\, P_3(x,x';x'')
\nonumber \\
&+i{\cal D}^2\av{\cpsi_g(x)\cpsi_g(x')\dpsi_g(x')\dpsi_g(x)} \left[
g(x')\hat{a}_F +h(x')\right]\, ,
\label{eq:cavitydP2dtlowint}
\end{align}
and so depends in turn upon a three-body correlation
\beq
P_3(x,x';x'') \equiv
\av{\cpsi_g(x)\cpsi_g(x')\Pop^+(x'')\dpsi_g(x')\dpsi_g(x)}\,.
\eeq
This continues up to the $N$th order -- leading to a hierarchy of equations of
motion for differing orders of correlations, which we give explicitly in
App.~\ref{app:cavityhierarchy}.   Similar hierarchies, but with greater
complexity, are encountered outside of the low intensity limit or in the case of
multilevel atoms~\cite{Ruostekoski1997a}.

In Eq.~\eqref{eq:cavitydP2dtlowint} the term proportional $P_2(x';x)$ on the right hand side represents recurrent scattering processes where an excitation between the atoms at the positions
$x$ and $x'$ is repeatedly exchanged by the scattering of the same photon. In a high-finesse cavity even a single atom can undergo recurrent scattering by repeatedly absorbing and emitting
the same photon. In cavity quantum electrodynamics this is referred to as the cooperative regime~\cite{CarmichaelVol2}. In the case of interaction of light with atomic ensembles in free space,
the cooperative terminology for scattering has traditionally had a notably broader meaning~\cite{guerin16_review}.

It rapidly becomes prohibitive to tackle the full hierarchy, and instead the
problem can be approached by a perturbative expansion, for example
in the small parameter $\rhoB/k^3$, which is valid at low atom
densities~\cite{Morice1995a,Ruostekoski1999a,vantiggelen90,bcs1}.  Rather than
such a perturbative approach, the field-theoretical model can be mapped onto stochastic electrodynamics
that is reminiscent of stochastic Langevin equation approach to a diffusion equation~\cite{Javanainen1999a,Lee16}.
The formalism of Ref.~\cite{Lee16} was derived for free space light propagation -- presenting techniques which exactly solve the
free-space counterparts of Eqs.~\eqref{eq:cavityfullopticalresponse} for the
case of atoms with a single ground level in the limit of low light intensity,
and which provide approximate but numerically practical solutions outside of
this limit.  Analogous approaches can also be introduced for light propagation in confined 1D waveguides~\cite{Ruostekoski_waveguide}.
Here we derive the stochastic electrodynamics for a cavity system in a semiclassical approximation.
We consider explicitly only the case of two-level atoms,
however the multi-level generalization follows from
Eqs.~\eqref{eq:cavityfullopticalresponse} and the discussions in
Ref.~\cite{Lee16}.

\section{Stochastic electrodynamics simulations of two-level atoms}
\label{sec:cavitytwolevel}

\subsection{Stochastic sampling}

Taking expectation values of Eqs.~\eqref{eq:cavitydPdt},
and~\eqref{eq:cavitydrhogdt} leads to a hierarchy of coupled equations of motion
for correlation functions describing the optical response of two-level atoms in
a cavity.  Computationally efficient solutions, particularly when spatial
variation of the atom density, cavity mode and detunings are nontrivial, are
provided by stochastic simulations of the optical response:  A set of $N$
discrete atomic positions $\{\stochx_1,\ldots,\stochx_N\}$ is sampled from the
thermal equilibrium $N$-body joint probability distribution function for the
atoms.  For each such realization of atomic positions, we solve the optical
response for the hypothetical model of $N$ atoms pinned to the fixed positions
$\{\stochx_1,\ldots,\stochx_N\}$.  Subsequent ensemble averaging over many such
independent realizations gives the quantum expectation values for the quantities
governing the optical response of the ensemble.

The appropriate $N$-body distribution governing the position sampling is that
representing the thermal equilibrium state of the atoms in the absence of light,
prior to the introduction of any pump beams into the cavity.  Since we have assumed
that the atoms are stationary, this distribution function for the discrete position
sampling is invariant.  The subsequent introduction of light can induce a
collective optical response, as governed by the equations of
motion~\eqref{eq:cavitydPdt}, and~\eqref{eq:cavitydrhogdt}.  These equations can be
integrated, or solved in the steady-state, for each single realization of discrete
positions.  The optical response of the ensemble, including the effects of quantum
statistical density correlations and light-induced atomic correlations, results
from the subsequent averaging over many such realizations.

In principle, any density correlations can be included in the distribution
function from which positions are drawn, for example Fermi-Dirac statistics can
be modeled using a Metropolis algorithm~\cite{Javanainen1999a,Ruostekoski_waveguide}.  However many
situations of interest can be sampled more straightforwardly -- for an
uncorrelated ensemble of classical atoms or an ideal Bose-Einstein condensate (BEC) the sampling
reduces to independently sampling the position of atom $i$ from the ground state
atomic density distribution of the ensemble in the absence of light.
Since we here concentrate on stationary atoms whose center-of-mass position is constant and only
the internal electronic state of the atoms evolves as a function of time,
the total one-body density $\rhoC_1(x)$ is given by this is initial ground-state atom density at all times.

To carry out this procedure we first need to find the equations of motion for the
atoms and light for each stochastic realization of atomic positions, which can be
obtained from the expectation values of Eqs.~\eqref{eq:cavitydPdt},
and~\eqref{eq:cavitydrhogdt} conditioned upon the set of positions
$\{\stochx_1,\ldots,\stochx_N\}$.  The conditioned expectation values of one-body
operators can be written in terms of one-body internal-level density matrix
elements 
\begin{align}
\av{\cpsi_e\dpsi_e}_{\{\stochxv_1,\ldots,\stochxv_N\}} &= \sum_j
\rhoD_{ee}^{(j)}\delta(x-\stochx_j), \label{eq:twolevelrhoedef}\\
\av{\cpsi_g\dpsi_g}_{\{\stochxv_1,\ldots,\stochxv_N\}} &= \sum_j
\rhoD_{gg}^{(j)}\delta(x-\stochx_j),\label{eq:twolevelrhogdef} \\
\av{\Pop^{+}(x)}_{\{\stochxv_1,\ldots,\stochxv_N\}} &= \Dc\pol\sum_j
\rhoD_{ge}^{(j)}\delta(x-\stochx_j)\,,\label{eq:sattwolevelPplusdef}
\end{align}
where $\rho_{ge}$ is the
matrix element $\langle g |\hat{\rho}|e\rangle$ which is related to $\langle
\hat{\psi}_g^\dagger\hat{\psi}_e\rangle$.

Eqs.~\eqref{eq:cavitydPdt}, and~\eqref{eq:cavitydrhogdt} then lead to the coupled
set of $2N$ nonlinear equations for a single realization of atomic positions
\begin{subequations}
\begin{align}
{d\over dt}\,&\rhoD_{ge}^{(j)} = \left(i\bar{\Delta}- \frac{{\cal
D}^2}{\kappabar}|g(\stochx_j)|^2\right)\rhoD_{ge}^{(j)}\nonumber \\
&+\frac{{\cal D}^2}{\kappabar}\sum_{l\neq j}
G_c(\stochx_j,\stochx_l)\left(2\rhoD_{ee}^{(j)}-1\right)\rhoD_{ge}^{(l)}
 \nonumber \\
&-i{\cal D}\left(2\rhoD_{ee}^{(j)}-1\right) \left[ g(\stochx_j)a_F +h(\stochx_j)\right]\,,
\label{eq:cavitystochasticP}\\
{d\over dt}\,&\rhoD_{ee}^{(j)} =
-2{\cal D}^2\frac{\kappa}{\kappa^2+\Delta_c^2}|g(\stochx_j)|^2\rhoD_{ee}^{(j)} \nonumber \\
&-2\mbox{Re}\left[\frac{\Dc^2}{\kappabar}\sum_{l\neq
j}G_c(\stochx_j,\stochx_l)\rhoD_{eg}^{(j)}\rhoD_{ge}^{(l)}\right]
\nonumber \\
&+2\mbox{Re}\left[i\Dc g(\stochx_j)\rhoD_{eg}^{(j)}a_F\right] \nonumber \\
&-2\mbox{Re}\left[i\Dc h^*(\stochx_j)\rhoD_{ge}^{(j)}\right]\,.\label{eq:cavitystochasticrhoe}
\end{align}
\label{eq:cavitystochastictwolevel}
\end{subequations}
Here we have used a
semiclassical factorization
approximation for all conditioned expectation values of
two-body correlation functions, e.g.,
\begin{align}
\av{\cpsi_e(x)&\hat{P}^{+}(x')\dpsi_e(x)}_{\{\stochx_1,\ldots,\stochx_N\}}\nonumber
\\ &=\Dc\sum_{\substack{j,l=1 \\ j \neq l}}^N\rhoD_{ee;ge}^{(j,l)}
\delta(x-\stochx_j)\delta(x'-\stochx_l)\,,\nonumber
\\
\rhoD_{ee;ge}^{(j,l)} &\simeq
\rhoD_{ee}^{(j)}\rhoD_{ge}^{(l)}\,.
\label{eq:cavityfactorization}
\end{align}
Such an approximation means that the stochastic equations are not able
to fully reproduce all the light-induced correlations in the system. However, the semiclassical approximation
provides the means to include saturation effects in our stochastic electrodynamics while avoiding the need to solve the full hierarchy of equations of
motion for quantum correlation functions discussed in the previous section.
Furthermore, the semiclassical approximation does not neglect all correlations,
since light-induced correlations that depend on the spatial positions of the
atoms are included.  In Sec.~\ref{sec:numericaltwoatomscavity} we compare in
detail the effect of the semiclassical approximation to the full quantum
field-theoretical solution for the case of two atoms in a cavity, and show that
the semiclassical approximation to the stochastic electrodynamics can still describe strong
cooperative effects.

Ensemble averaging the single realization results from
Eqs.~\eqref{eq:cavitystochastictwolevel} over many stochastic realizations of
atom positions allows the calculation of physical observables. We could also extend the stochastic electrodynamics to include quantum effects beyond the semiclassical factorization of Eq.~\eqref{eq:cavityfactorization}. This constitutes of including
the correlation functions between the different internal atomic levels, as described in Ref.~\cite{Lee16} for multilevel atoms in free space. In the lowest order such correlation functions include
the pair correlations between the internal levels of all atom pairs.

We note that in a
cavity, the cavity field itself may be a significant contribution to the
confining potential for the atoms, in addition to any external potential.  In
such a situation a self-consistent solution must be found, such that the cavity
amplitude reflects the assumed $N$-body distribution function from which the
atomic positions are sampled.

\subsection{Limit of low light intensity}

A limit which is both useful and important conceptually is that of low light
intensity.
In this limit, dropping terms of higher than first order
in the light field amplitude or excited state operators leads to the simpler
linear equation for the atomic ensemble in a single stochastic realization of atomic positions
\begin{align}
{d\over dt}\,&\rhoD_{ge}^{(j)} = \left(i\bar{\Delta}- \frac{{\cal
D}^2}{\kappabar}|g(\stochx_j)|^2\right)\rhoD_{ge}^{(j)}\nonumber \\
&-\frac{{\cal D}^2}{\kappabar}\sum_{l\neq j}
G_c(\stochx_j,\stochx_l)\rhoD_{ge}^{(l)}
 \nonumber \\
&+i{\cal D} \left[ g(\stochx_j)a_F +h(\stochx_j)\right].
\label{eq:cavitydiscretelowlight}
\end{align}
To first order in light field amplitude $\av{\cpsi_{e}\dpsi_{e}}$ vanishes, and
$\av{\cpsi_{g}\dpsi_{g}}$ is invariant.  Each stochastic realization in this
limit is equivalent to solving the response of a  model of linear harmonic
oscillators at positions $\stochx_1,\ldots,\stochx_N$ within an optical cavity.
The importance of the limit of low light intensity is that, in contrast to
the case where saturation is non-negligible, for atoms with a single relevant
electronic ground level the stochastic solutions can be shown to fully reproduce
the dynamics of the full hierarchy of correlation functions~\footnote{This can
be shown by arguments that are analogous to those presented in App.~B
of Ref.~\cite{Lee16}, starting from \eqref{eq:cavitylowintensityhierarchy}.}, and so in
this limit the accuracy of the stochastic method is limited only by the
statistics of the number of realizations that are generated.
In the following section we calculate the eigenmodes of the cavity-atom
system, analyzing the limit of low light intensity in detail.

The stochastic electrodynamics equations \eqref{eq:cavitydiscretelowlight} or \eqref{eq:cavitystochastictwolevel} describe the optical response
of an atomic ensemble in a cavity. The initial distribution of the atoms and their quantum statistics before the light enters the system can be sampled
from the appropriate joint probability distribution of atomic positions, provided that such a distribution can be synthesized~\cite{Javanainen1999a}. The ensemble-averaging
the time evolution is then reminiscent of the stochastic Langevin equation solution to the diffusion equation. Many-atom cavity responses can also be simulated using
stochastic phase-space methods based on the Wigner representation of the atomic and light fields~\cite{Lee2014a,Lee2015a}. Here the initial quantum distribution of the
atoms is also represented by stochastic sampling in the corresponding Langevin equation, obtained from the appropriate phase-space distribution.  Each stochastic realization of the atomic wavefunction is then evolved
according to the dynamical equations that also include a stochastic Wiener noise increment, resulting from the dynamics that is conditioned on the leaking of photons
out of the cavity. However, in the phase-space approach the emphasis is on the center-of-mass motion of the atoms, and the atomic polarization is treated as a continuous field
-- an approximation that generally neglects the position-dependent light-induced correlations that we investigate in this paper.

\section{Eigenmodes and optical response of the cavity system}
\label{sec:eigenmodes}

\subsection{Eigenmodes in the limit of low light intensity}

\subsubsection{Structure of the eigenmodes}

In the limit of low light intensity, the equation of motion for the polarization
density [Eq.~\eqref{eq:cavitydiscretelowlight}] for a single realization of
discrete atomic positions can be written in the form
\beq
\frac{d}{dt}\rhoD^{(j)}_{ge} = \sum_{l=1}^N{\cal
M}_{jl}\rhoD^{(l)}_{ge}+F^{(j)}
\eeq
where $F$ involves only the driving field terms.  The matrix ${\cal M}$ therefore
accounts for the cavity mediated interparticle interactions, along with the
single particle cavity evolution and self-interaction of the atom with the cavity
field.  The eigenmodes of ${\cal M}$ are then instrumental in understanding the
response of the atomic ensemble at a given cavity-pump detuning. In general, we find $N$ eigenmodes, however only
one of the $N$ eigenvectors is certain to have a non-zero overlap with the cavity
mode $g(x)$ and therefore to couple to the cavity mode.  This mode involves all
atoms collectively polarized with directions determined by the local sign of
$g(x)$, and represents the strongest possible collective coupling of atoms to the cavity
mode.  It is characterized by a decay rate (linewidth) of the order of
$Ng_0^2/\mbox{Re}[\kappabar]$ and frequency (line shift) $Ng_0^2/\mbox{Im}[\kappabar]$, it
therefore decays much faster than would be expected for a single atom, and
represents a superradiant mode.

In contrast, the other $N-1$ modes are subradiant, characterized by much narrower
linewidths. In fact, in the absence of some non-periodic dependence on $g(x)$ generally
the subradiant modes are completely orthogonal to the cavity mode and hence are
infinitely long lived (trivially, since we neglect in our model any loss other than to
the cavity mode).  The exception to this statement occurs when the atomic detuning
$\bar{\Delta}(x)$ is spatially varying. Such a detuning is generally introduced by a
trapping potential and can be tailored in practice by employing a magnetic field strength
gradient. Since ${\cal M}$ includes $\bar{\Delta}(x)$, any change in the detuning
perturbs the eigenmodes of the system.  If $\bar{\Delta}(x)$ is not uniform or of a form
proportional to $g(x)$, then the perturbation of the eigenmodes is not merely a trivial
frequency shift. Instead it results in the coupling of subradiant modes to the cavity
mode and therefore in non-zero decay rates for the subradiant modes.  Consequently, in
subradiant mode results presented in this paper we necessarily re-diagonalize ${\cal M}$
for each different $\bar{\Delta}(x)$ considered. Of course, realistic systems will always
have some subradiant mode decay even in the absence of coupling to the cavity due to loss
to free space modes transverse to the cavity.  We do not include any such loss mechanism
in our model, which allows easy identification of decay to the cavity mode, and
subsequent measurement via the cavity output.  The limit of weak transverse decay may be
approached, for example,  by trapping the atoms in lattices with
subwavelength-spacing~\cite{Sutherland1D,Bettles1D}.

In addition to the collective atomic superradiant and subradiant
modes, the cavity mode itself has a resonance (at $\Delta_c=0$ in the absence of
atoms).  The full cavity-atom system therefore admits $N+1$ normal modes, whose
nature may be mixed.  In particular, the superradiant atomic and cavity modes
exhibit an avoided crossing near $\bar{\Delta}=\Delta_c= 0$, leading to the
characteristic ``vacuum Rabi
splitting''\cite{SanchezMondragon1983,CarmichaelVol2} behavior illustrated in
Fig.~\ref{fig:DressedStatesScan}.  In contrast to the superradiant mode, the
subradiant modes remain unperturbed by the cavity mode frequency.  In the
remainder of the paper, when considering collective atomic modes we will
generally work in the weak-coupling regime where the cavity mode is far-detuned
from resonance such that the atomic modes can be considered in isolation.

\subsubsection{Lattice system}

As an example, we consider a small system of $8$ atoms confined in a lattice
commensurate with the cavity mode, such that atoms are confined near maxima of
$|g(x)|$, and we will assume that the atoms are in a Mott insulator (MI)
state~\cite{Morsch06}
of exactly one atom per lattice site.  In practice, a harmonic potential is
often used in conjunction with an optical lattice and leads to a ``wedding
cake'' MI ground-state of differing occupancies. A system of one atom per site
can then be achieved
by manipulating the sites with excess occupancy \cite{singlespin}.
The quantum phase of the atomic ensemble
can be important in determining the optical response, and it plays a central
part in our stochastic method since it determines the joint-probability
distribution from which single realizations of atomic positions are sampled.
For an MI state with one atom per site the stochastic sampling for each site is similar to that for independent atoms~\cite{Jenkins2012a,Lee16}:   We take the atoms to be confined along their one degree of
freedom ($x$) by an external lattice potential commensurate
with the cavity mode, giving rise to Wannier functions $\phi_i(x) =
\phi(x-\ell_i)$ centered at antinodes $\ell_i$ of the
coupling strength $g(x) = g_0 \sin(kx)$ in a Fabry-Perot cavity.  Here
$\phi(x)$ is given by
\begin{equation}
  \label{eq:Wanier1}
  \phi(x) \simeq \frac{1}{(\pi L_x^2)^{1/4}}
  \exp\left(-\frac{x^2}{2 L_x^2}\right) \text{,}
\end{equation}
where $L_x$ governs the confinement of the well, and the linear density of a
single atom in site $i$ is $\rhoE{i}{x}$.
A single realization of discrete atomic positions for the stochastic
method of Sec.~\ref{sec:cavitytwolevel} is then found by sampling a single position
from the distribution $\rhoE{i}{x}$ for each occupied lattice site.  Later, in
Sec.~\ref{sec:manybody} we will contrast these results with those for atoms in
an ideal superfluid state.

In order to couple the subradiant atomic modes to the cavity optical response, we
apply a simple linear gradient to the detuning, such that $\bar{\Delta}(x) =
\bar{\Delta}_0+\bar{\Delta}_1 x/\lambda$. Ensemble averaging the eigenmodes found
from many stochastic realizations of discrete atomic positions leads to the
distributions of eigenmode decay rates and frequencies shown in
Fig.~\ref{fig:N8eigenmodes}.  A single superradiant mode appears, whose
distribution of decay rates and frequencies mirrors closely the real and imaginary
parts, respectively, of the distribution of $\sum_j g^2(X_j)/\kappabar$.  In
addition, there are also $7$ non-trivial subradiant modes with linewidths three
orders of magnitude smaller than the superradiant mode.  In this example the
subradiant modes are conveniently separable in frequency, but this is not
generally the case. The linear detuning gradient used here is responsible for the
uniform frequency spacing of the subradiant modes, and increasing the number of
atoms in the system will lead to additional subradient modes with the same
spacing.  

\subsection{Optical response in the low light intensity limit}

In the previous section we illustrated the calculation of the many-atom collective
eigenmodes in the cavity system, where the atomic detuning is subject to a linear
spatial variation that couples the sub-radiant eigenmodes to the cavity mode.
Next we show that these modes, when coupled in this manner, can be excited and
identified in the optical response to incident light.  We then show that these
modes can also be preferentially excited by suitable transverse light.

When coupled to the cavity mode, each of the collective modes
can be excited by axially pumping the cavity on resonance with the corresponding
mode frequency.  Figure~\ref{fig:N8moderesonances} demonstrates the resonances
in the steady-state cavity intensity $|\av{\dopa}|^2$, directly proportional to
the transmitted light intensity, obtained from an ensemble of stochastic
realizations of Eq.~\eqref{eq:cavitydiscretelowlight} for each value of the
axial pump frequency. A full scan of frequencies for a given atom-cavity
detuning shows a broad cavity mode resonance, with a linewidth of approximately
$\kappa$, a superradiant mode with a linewidth of $\sim
Ng_0^2/\mbox{Im}[\kappabar]$, and the very narrow and comparatively weakly
excited subradiant modes.   

\begin{figure}
\includegraphics[width=0.95\columnwidth]{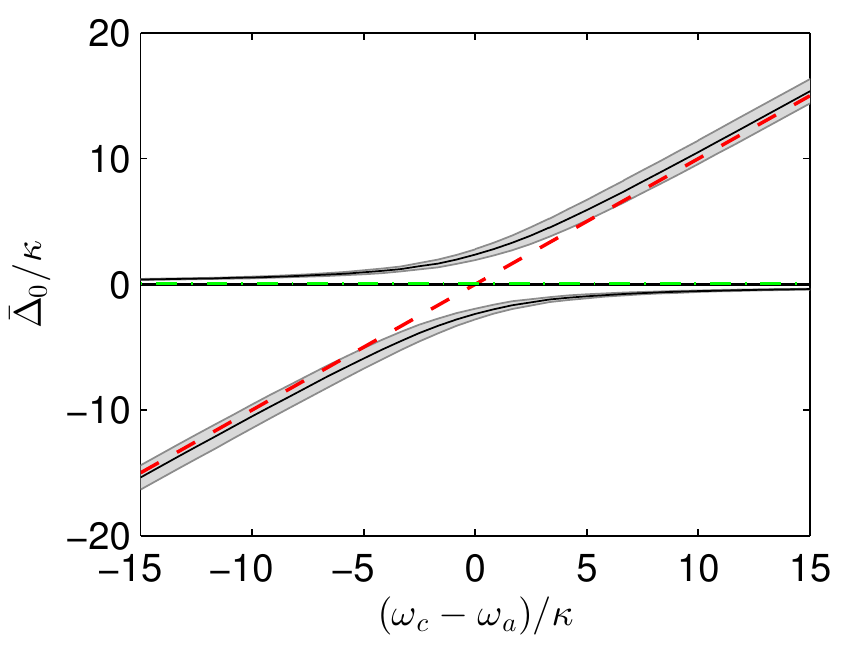}
\caption{ Normal modes of the cavity-atom system for a system of
8 atoms confined by an optical lattice potential and in a MI state
of one atom per site.  The bare
superradiant atomic mode (green, dot-dashed line) and cavity mode (red, dashed
line) are dressed by the atom-light interaction to form the dressed states
which exhibit an avoided crossing (solid black lines, shading represents the
linewidths of the associated mode).  In contrast, the subradiant modes remain at
fixed $\bar{\Delta}_0$ for all $\omega_c$, their position is represented by the
horizontal black line, as they cannot be resolved on this scale.  System parameters for this
simulation are $g_0/\kappa = 0.9$, $\bar{\Delta}_1/\kappa =
1.4\times 10^{-3}$, $\kappa = 342\omega_R$ where $\omega_R$ is the cavity mode
recoil frequency.  The atoms are confined about each lattice site $\ell_i$ with a
density distribution $\rhoE{i}{x} \propto \exp\{-[(x-\ell_i)/(0.08\lambda)]^2\}$.
\label{fig:DressedStatesScan}}
\end{figure}

\begin{figure}
\includegraphics[width=0.49\columnwidth]{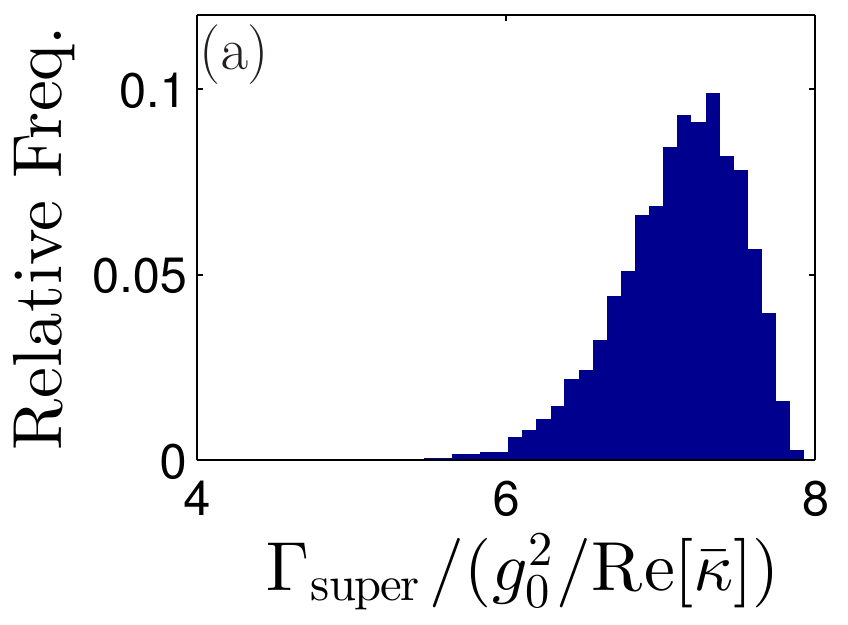}
\includegraphics[width=0.49\columnwidth]{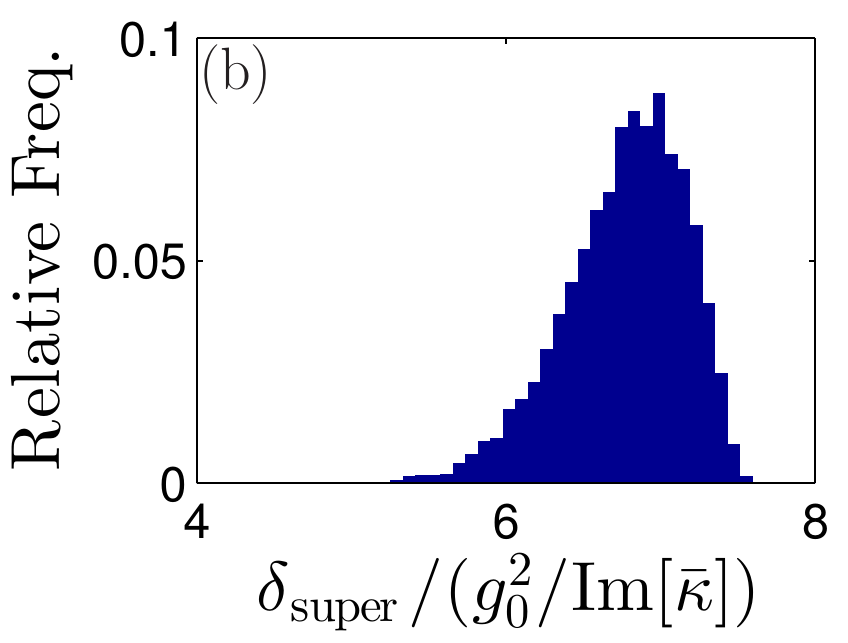}\\
\includegraphics[width=0.49\columnwidth]{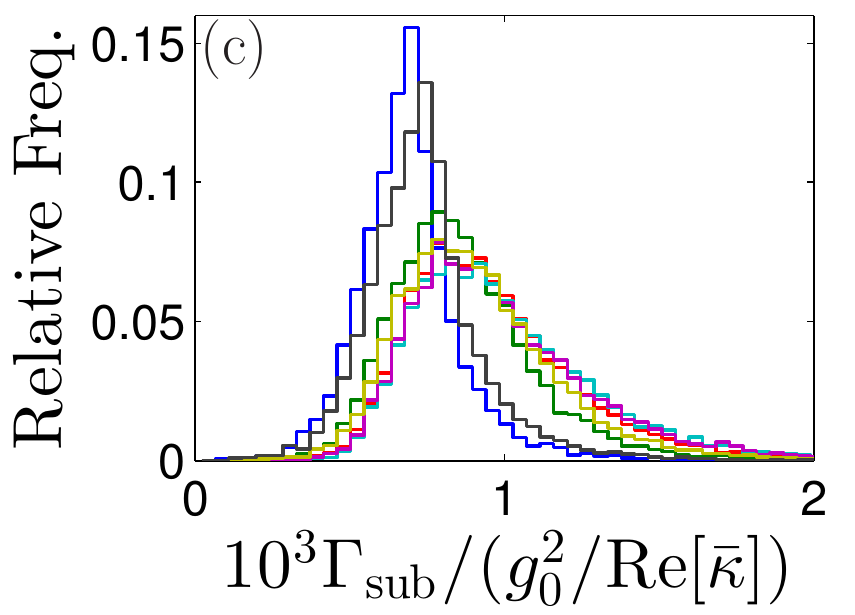}
\includegraphics[width=0.49\columnwidth]{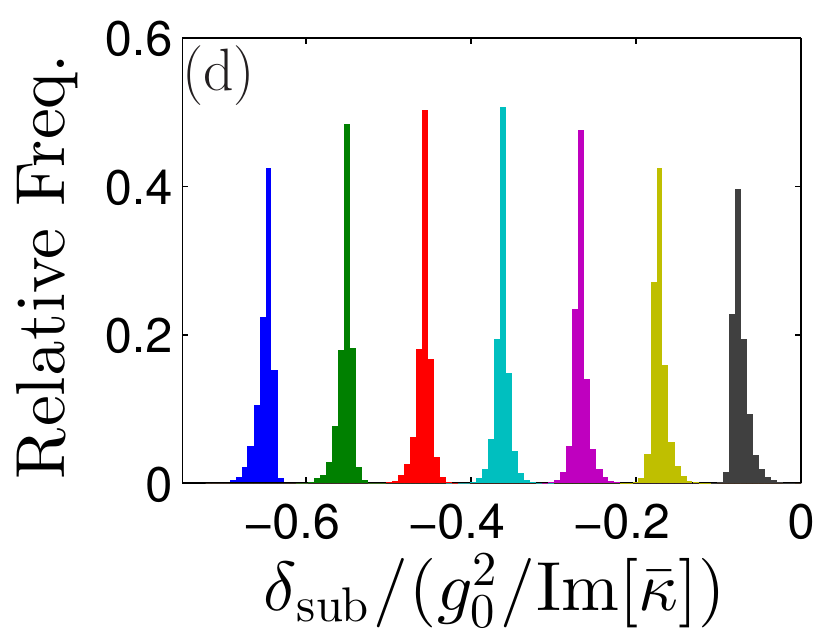}
\caption{Distribution of eigenmode decay rates $\Gamma$ and
frequencies $\delta$, for an ensemble of individual realizations of stochastic
atomic positions.  The superradiant mode is shown in (a) and (b) while the $7$
subradiant modes are illustrated in (c) and (d).  System parameters as for
Fig.~\ref{fig:DressedStatesScan}, with $\Delta_{cp} = -100\kappa$ and $\bar{\Delta}_0 =
0$ and a linear gradient of the detuning
 $\bar \Delta_1/\kappa = 1.4\times 10^{-3}$. \label{fig:N8eigenmodes}}
\end{figure}

\begin{figure*}
\includegraphics[width=0.49\columnwidth]{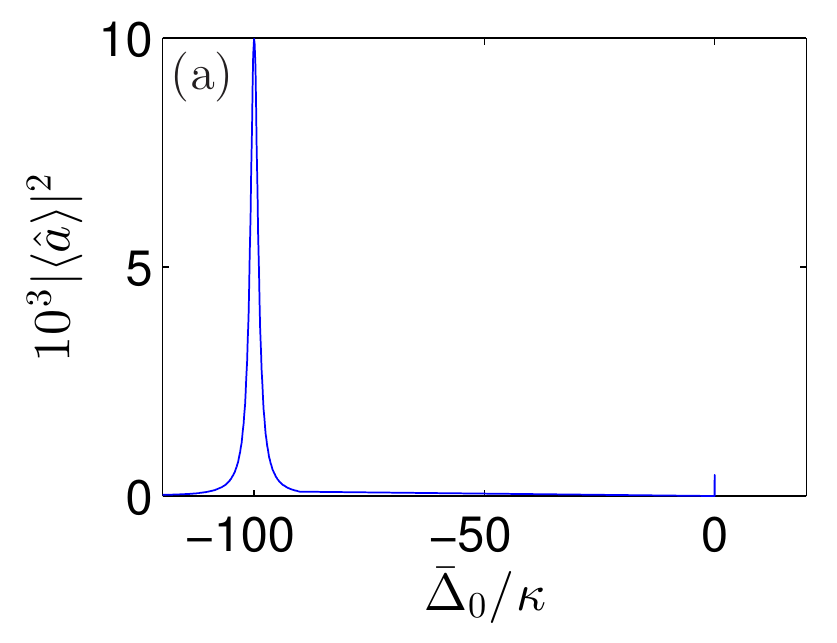}
\includegraphics[width=0.49\columnwidth]{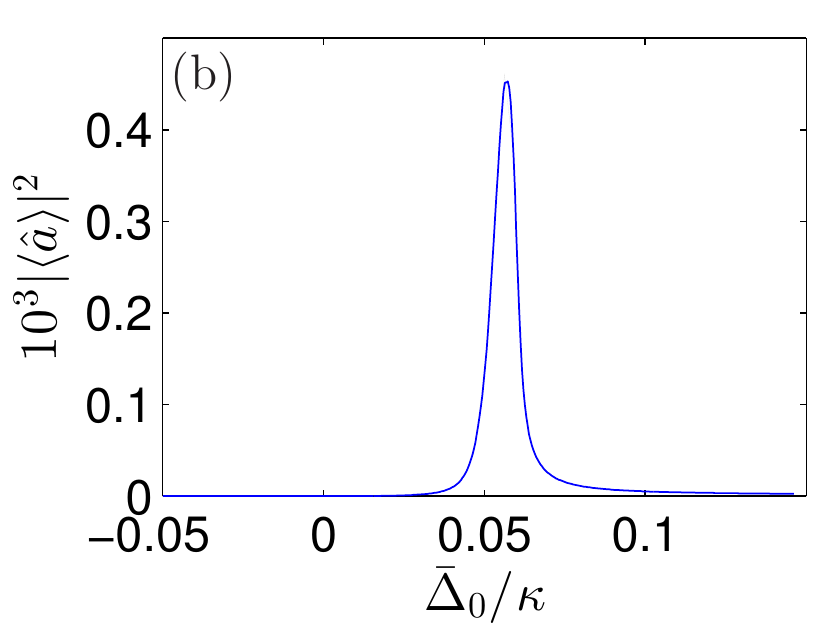}
\includegraphics[width=0.49\columnwidth]{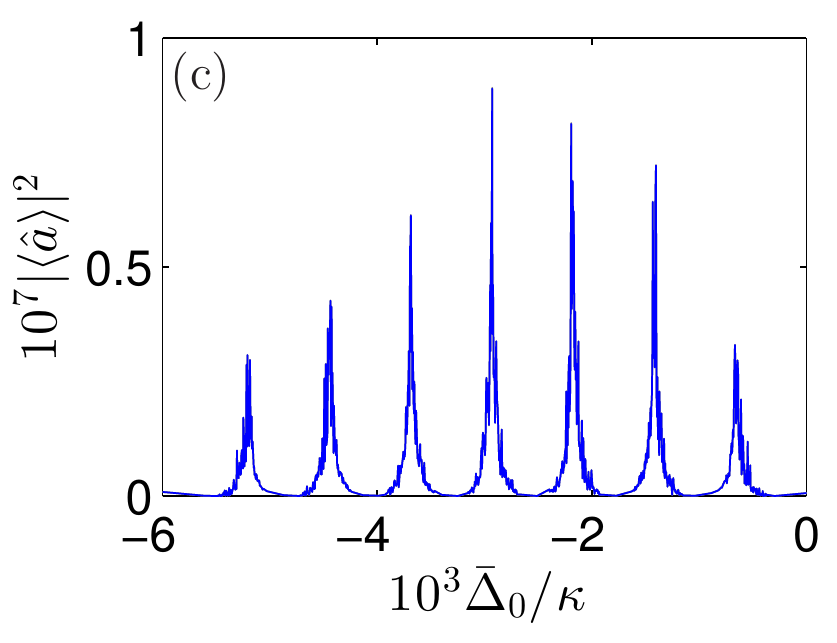}
\caption{ Spectra for the optical response of the many-atom cavity system.
Resonance peaks in the steady state cavity light intensity $|\av{\dopa}|^2$ due
to the resonance with (a) cavity mode [the superradiant mode resonance is also
resolvable near $\bar{\Delta}_0 =0$], (b) the superradiant mode, and (c) the
subradiant modes. Parameters as for Fig.~\ref{fig:DressedStatesScan}, but with
$\Delta_{ca} = \bar{\Delta}_0-\Delta_{cp}$ held constant at $-100\kappa$ while
the pump frequency $\bar{\Delta}_0$ is scanned.  Results calculated in the limit
of low light intensity. \label{fig:N8moderesonances}}
\end{figure*}

\begin{figure*}
\includegraphics[width=0.49\columnwidth]{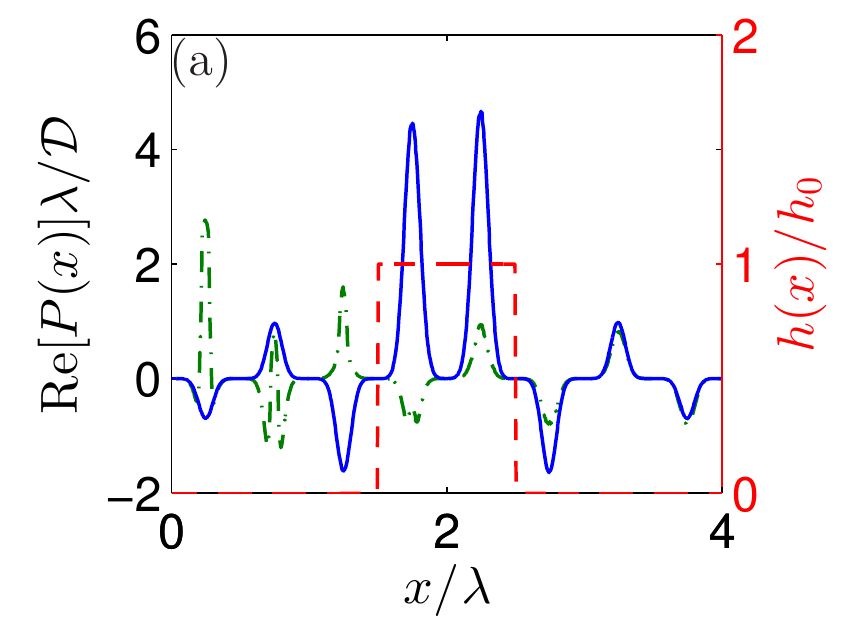}
\includegraphics[width=0.49\columnwidth]{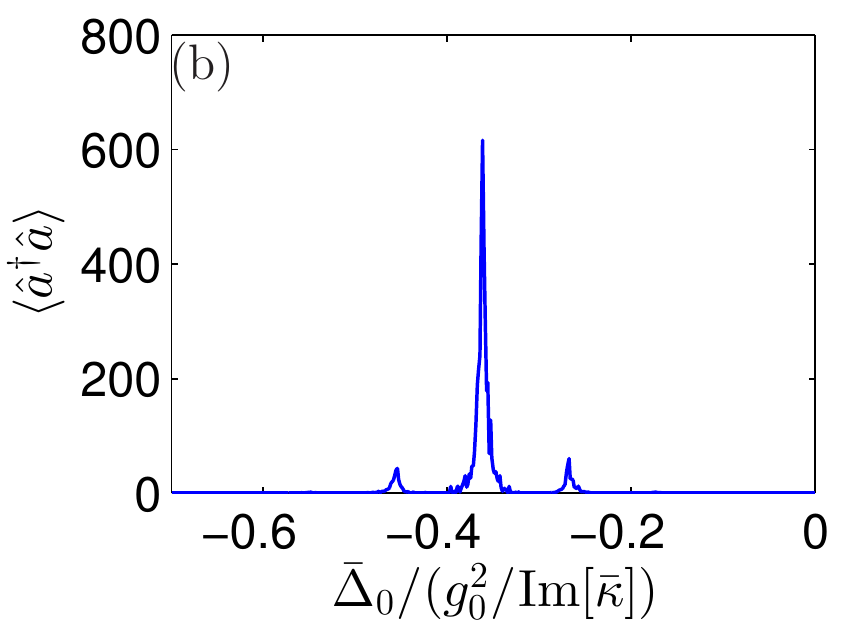}
\includegraphics[width=0.49\columnwidth]{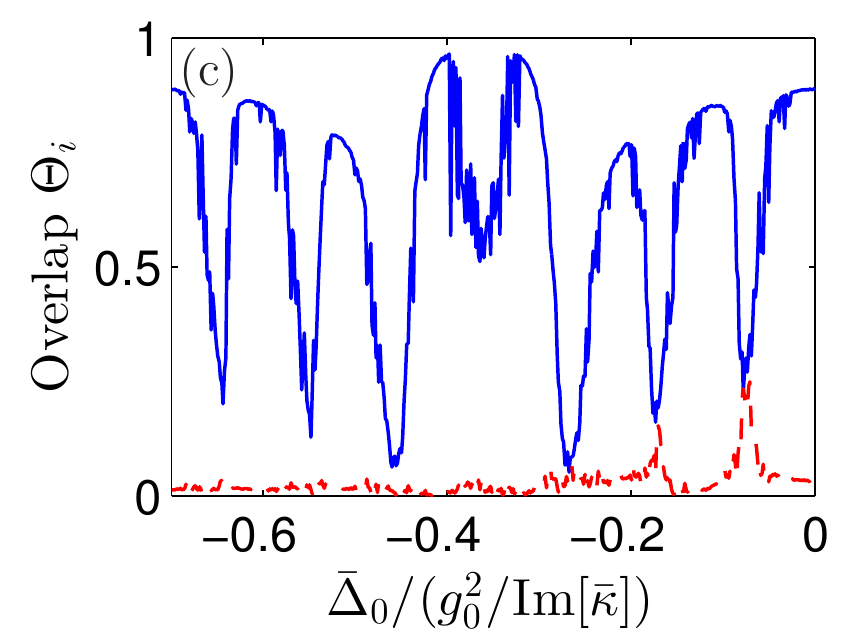}
\caption{(a) Subradiant mode spatial profile (blue, solid)
corresponding to the center-most mode in Fig.~\ref{fig:N8eigenmodes}(d) and
transverse pump profile $h(x)$ tailored to excite this mode (red, dashed). For
comparison, the spatial profile of the superradiant mode is also shown (green,
dot-dashed).  Note that the imaginary part of each profile is negligible. (b)
Response of the steady-state cavity light intensity to the tailored transverse pump beam as a function of
pump-atom detuning $\bar{\Delta}_0$.  (c)   Overlap of the excited polarization
density $P(x)$ with the spatial profile of a particular collective mode $P_i(x)$,
defined as $\Theta_i = \left|\int P(x)P_i^*(x) dx\right|/\sqrt{\int |P(x)|^2
dx\int |P_i(x)|^2 dx}$.  Overlaps are shown for the targeted subradiant mode
(blue, solid) and the superradiant mode (red, dashed), and clearly show a
predominant excitation of the desired subradiant
mode.\label{Fig:subr_transverseresponse}}
\end{figure*}

While the subradiant modes only couple weakly to the cavity mode, they can also
be excited by a suitably tailored transverse pump $h(x)$.  Under an ensemble
average of many realizations, the spatial profile of the centermost of the
subradiant modes in Fig.~\ref{fig:N8eigenmodes}(d) corresponding to the linearly varying
detuning can be determined, and is
shown in Fig.~\ref{Fig:subr_transverseresponse}(a).  In contrast to the
superradiant mode in which the polarizations of neighboring lattice sites are
out of phase, in this case the subradiant mode has the centermost sites in phase
(the exact distribution depends on the spatial form of $\bar{\Delta}(x)$ along
with the other spatial dependence in the system).  A simple tailoring of a
transverse beam $h(x)$ which might predominantly excite this mode is suggested by
the step function in Fig.~\ref{Fig:subr_transverseresponse}(a)\footnote{We
have used a rectangular transverse pump beam for clarity in demonstrating the
effect. In practice, a rectangular beam with width $\lambda$ may be difficult to
realize experimentally.  However, any reasonably narrow beam which predominantly
and symmetrically excites the central two lattice sites should suffice.}.  Using this
transverse beam as the only pump mechanism, and calculating the steady-state
cavity intensity as a function of pump frequency, we see
in Fig.~\ref{Fig:subr_transverseresponse}(b) that indeed the targeted (centermost) subradiant
mode is strongly excited at the appropriate resonance frequency.  The
neighboring subradiant mode also shows a weak response, but no other subradiant
modes are appreciably excited.  In order to demonstrate that the subradiant mode has indeed been excited,
in preference to the superradiant mode, Fig.~\ref{Fig:subr_transverseresponse}(c) shows the overlaps of the
excited polarization density $P(x)$ with the spatial profile of the superradiant and the targeted
subradiant modes.  Over the frequency range where the subradiant mode shows a strong response it can clearly
be seen that the superradiant mode is only negligibly excited.  The subradiant mode excited by this
tailored transverse pump beam has a decay rate distribution peaked at $8.5\times 10^{-4}
g_0^2/\mbox{Re}[\bar{\kappa}]$ [see Fig. 2(c)], which is almost four orders of
magnitude smaller than that of the superradiant mode.

\subsection{Eigenmodes in the saturated case}

Outside of the limit of low light intensity, where it is necessary to include the
effects of saturation, the description of the optical response becomes more
complicated.  We must consider the coupled system of $3N$ nonlinear equations of
motion \eqref{eq:cavitystochastictwolevel} for $\rhoD_{ge}^{(j)}$, $\rhoD_{eg}^{(j)} =
\rhoD_{ge}^{(j)*}$ and $\rhoD_{ee}^{(j)}$.  Given a steady-state solution ${\bf v}_0$ of
Eqs.~\eqref{eq:cavitystochastictwolevel}, the response to a small perturbations
${\bf\Delta v}$ about that point are encoded in the Jacobian matrix
${\bf J}({\bf v}_0)$ evaluated at the steady state
\beq
\frac{d{\bf\Delta v}}{dt} = {\bf J}({\bf v}_0) {\bf\Delta v} .
\eeq
The eigenmodes of ${\bf J}({\bf v}_0)$ can be useful to describe the behavior
of the system near this steady state. We have $3N$ eigenmodes of the atom
system, which must then be coupled to the eigenmode of the cavity. At low light
intensities, $\rhoD_{ee}^{(j)}$ plays a negligible role in the dynamics, and the
eigenmodes can be grouped into sets:  $N$ conjugate pairs, which are eigenmodes
predominantly representing response of $\rhoD_{ge}^{(j)}$ and its conjugate,
which tend in the limit of low intensity to those discussed in the previous
section.  There are then $N$ eigenmodes which describe the response of the
(almost negligible) $\rhoD_{ee}^{(j)}$. The latter modes have eigenvalues
predominantly determined by the values of $g^2(X_j)/ \kappabar$ for a single
realization, and so simply represent the self interaction of an excited
state atom via the cavity mode at point $X_j$.

\section{Comparison of semiclassical approximation with full
treatment for two atoms}
\label{sec:numericaltwoatomscavity}

In this section we compare the stochastic results including saturation of
Sec.~\ref{sec:cavitytwolevel} with the results of the full quantum
field-theoretical representation of the optical response for coherent scattering
for the simple test case of two atoms.

In Sec.~\ref{sec:cavityformalism} we derived quantum field-theoretical equations
of motion governing the optical response of an ensemble of atoms in an optical
cavity.  We then discussed how these equations can be simulated by stochastic
electrodynamics simulations in Sec.~\ref{sec:cavitytwolevel}.  We noted that the
factorizations similar to \eqref{eq:cavityfactorization} in the derivation of
electrodynamics equations of motion introduce semiclassical approximations to
the calculations that go beyond the limit of low light intensity. Even in the limit
of low light intensity, some approximate factorizations are necessary to
describe atoms with multiple electronic ground levels~\cite{Lee16}.  Here we compare the results of the
stochastic simulations with the full field-theoretical solution of a coupled set
of equations for correlation functions of atomic densities and polarizations
that does not rely on the factorization approximation of
Eq.~\eqref{eq:cavityfactorization}. The full field-theoretical solution is exact
for atoms which are stationary regarding their center-of-mass motion.
In particular, we neglect the potential effect of the cavity field on the atomic recoil or
on the confining potential experienced by the atoms.

Equations (\ref{eq:cavitydPdt}) and~(\ref{eq:cavitydrhogdt}) show that single
particle correlation functions, such as the polarization and level populations,
in general depend on second-order correlation functions.  In turn, each of these
depends on higher-order correlation functions, leading to a hierarchy of
equations of motion which continue until the $N$th order.  Although this is a full quantum field-theoretical treatment for the system, it is not feasible to solve this
full hierarchy of coupled equations for large $N$, which motivates the method of
stochastic electrodynamics simulations discussed in this paper. However, it is feasible to analyze the field-theoretical treatment for the simple test case of a pair of two-level atoms, as
described in App.~\ref{app:cavityhierarchy}.  We are therefore able to
compare the semiclassical approximation to the stochastic electrodynamics
(Sec.~\ref{sec:cavitytwolevel}) with the exact solution.

Again, we take the atoms to be confined in an optical lattice, with linear
densities in each site governed by the Wannier functions of
Eq.~\eqref{eq:Wanier1} for $i={1,2}$. We assume a MI state of the
atoms, with a single atom in each site.  As explained in the previous section,
for the stochastic method the discrete particle positions are then sampled from
the corresponding joint probability distribution for the ground state densities
in the absence of cavity light
\begin{equation}
\rhoC_2(x,x') =
\frac{1}{2}\left[\rhoE{1}{x}\rhoE{2}{x'}+\rhoE{2}{x}\rhoE{1}{x'}\right].
\label{eq:Mottrho2}
\end{equation}
In the absence of cavity light, this expression must also be equal to
$\rhoG_2(x,x')$ in the atom number conservation condition given by
Eq.~\eqref{eq:cavityhierarchy_atomnumbercons}.  And since we have assumed that the
atoms are stationary with respect to their center of mass motion, this remains the
case even once light enters the cavity.   With this identification, the full
hierarchy of equations of motion for this system can be solved as shown in App. B.  Below, we compare the two
techniques for the different manifestations of collective optical response.

\subsection{Superradiant mode}

\begin{figure}
\includegraphics[width=0.49\columnwidth]{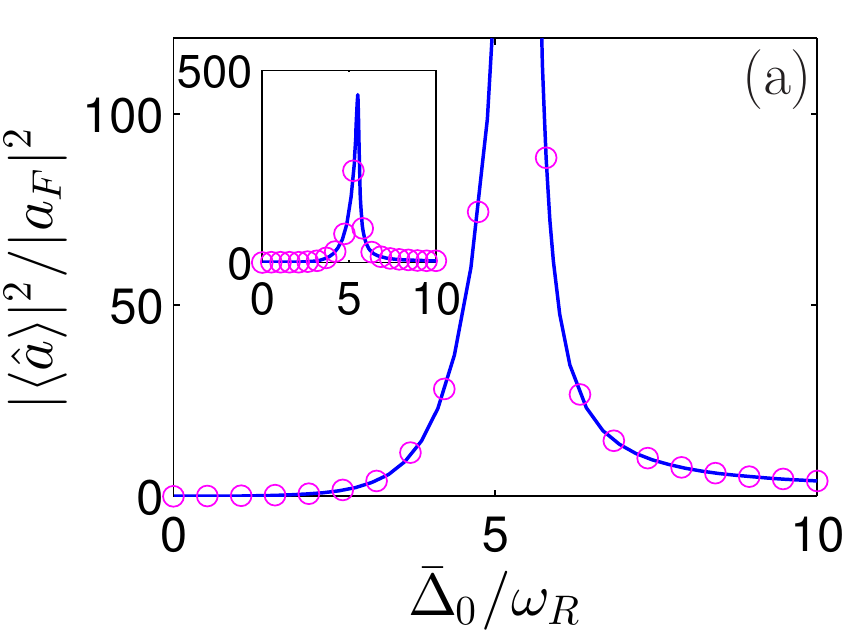}
\includegraphics[width=0.49\columnwidth]{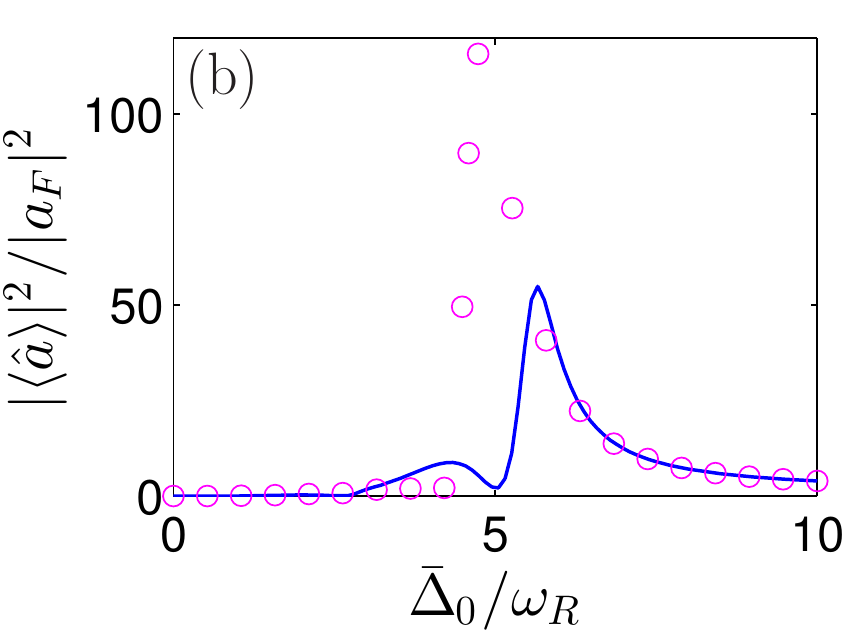} \\
\includegraphics[width=0.49\columnwidth]{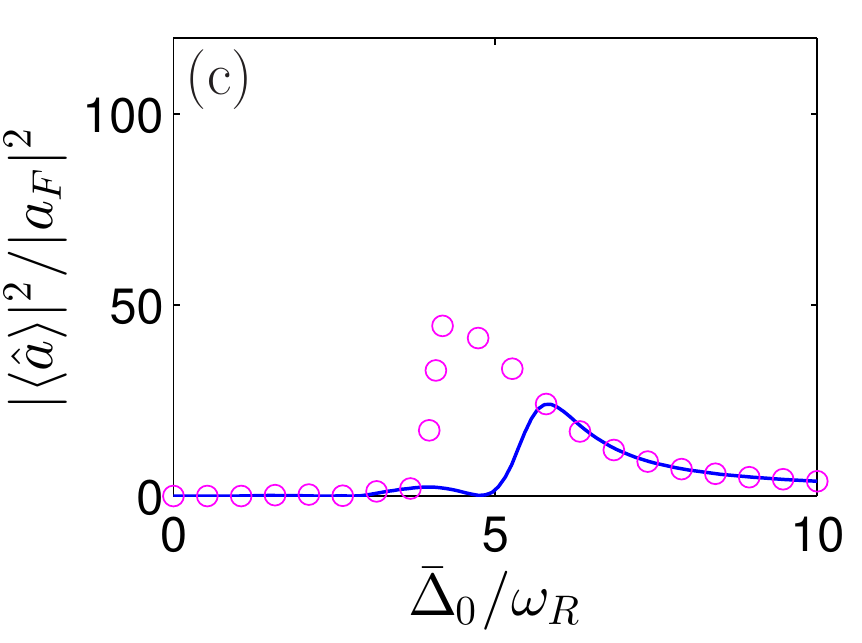}
\includegraphics[width=0.49\columnwidth]{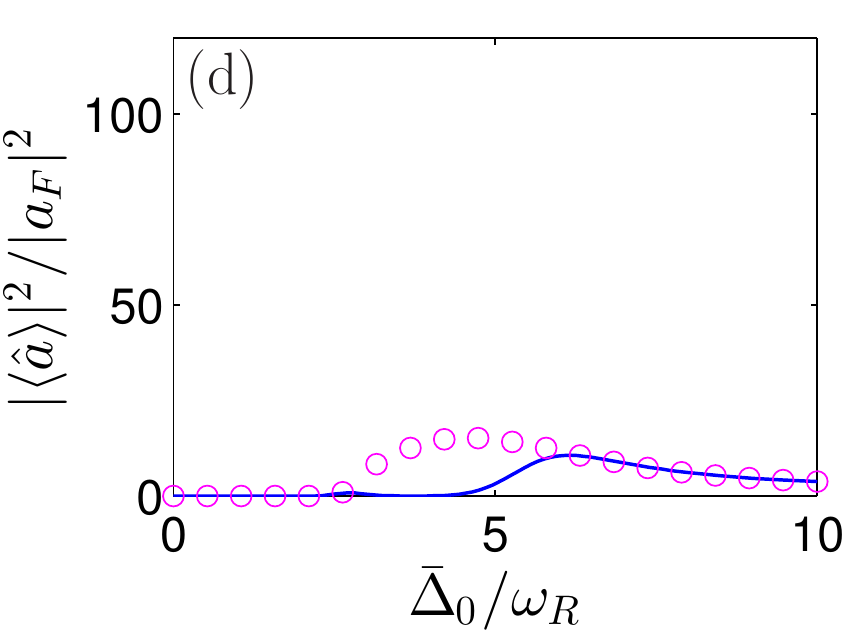} \\
\includegraphics[width=0.49\columnwidth]{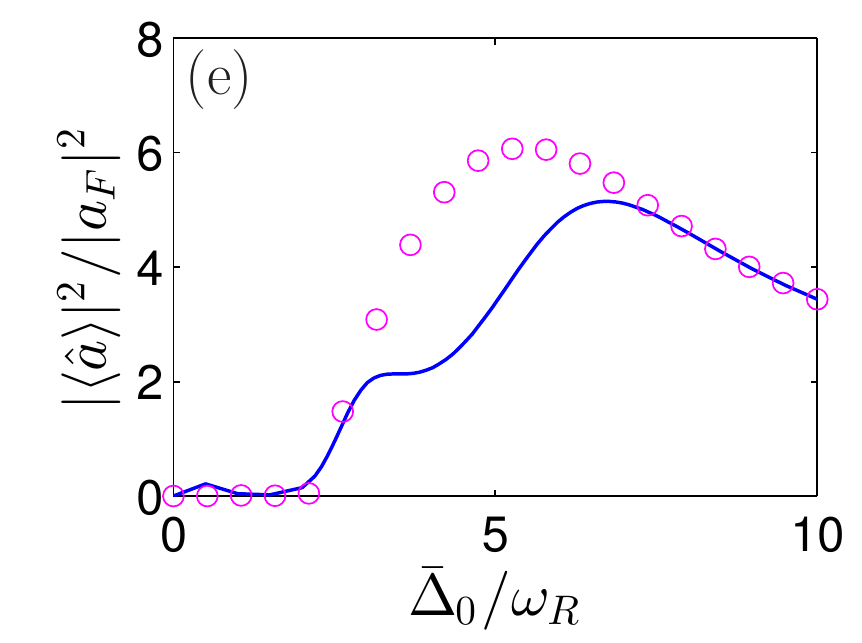}
\includegraphics[width=0.49\columnwidth]{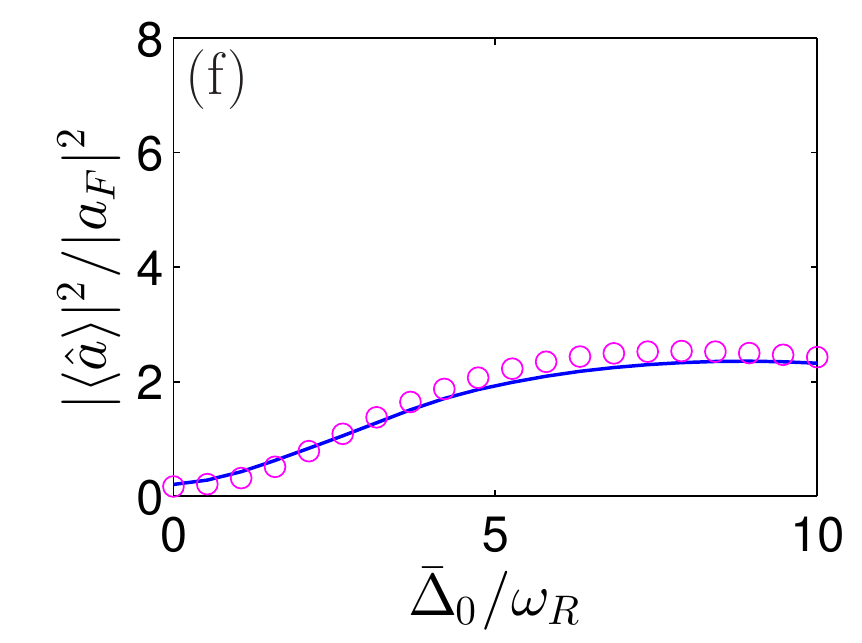}
\caption{The spectrum of the
steady-state light intensity inside the cavity for two atoms, showing a resonance in intensity due to
the superradiant mode for a two atom system, and including the effects of
saturation.  Results are shown for full quantum field-theoretical results
(lines) and semiclassical approximation to stochastic electrodynamics (circles) for 
the low intensity limit (a) [and showing the full extent of the resonance peak in the
inset], and for pump strengths $\eta$
of $0.05\kappa$ (b),  $0.1\kappa$ (c), $0.2\kappa$ (d), $0.4\kappa$ (e), and
$1\kappa$ (f) [note the change in scale for (e)-(f)].  For (b)-(f) the peak
saturations $\int \av{\cpsi_e(x)\dpsi_e(x)} dx/N$ reached for each intensity are,
respectively, $0.2$, $0.3$, $0.42$, $0.47$ and $0.49$.  In the low intensity limit the
superradiant mode has no significant structure, but once saturation is included a
two peak structure emerges in the full quantum results.  The semiclassical approximation is
unable capture all of this detail, although it does well on the large detuning
side of the resonance.  At higher intensities the profile broadens and the
structure is lost, at these intensities the semiclassical approximation agrees well with
the exact results.\label{fig:ExactvsStoc_superradiant}}
\end{figure}

As described in the previous section, a two-atom system admits two collective
atomic modes: the superradiant mode along with a single subradiant mode.
Figure~\ref{fig:ExactvsStoc_superradiant} shows the spectrum of the steady-state optical cavity response for the two-atom system
for different values of the pump strength (or atomic saturation). We display the light intensity inside the cavity close to resonance
with the superradiant eigenmode. The cavity is
far-detuned from the pump frequency so that the cavity and collective atomic
modes are only weakly coupled.  In the limit of low light intensity, the
resonance peak is Lorentzian, and we confirm that the stochastic electrodynamics
converges to the result obtained by solving the correlation functions.  At intermediate intensities, when the stochastic electrodynamics is only approximate, the
results become more complicated.  Firstly, the results from the full treatment of
correlation functions show that structure appears in the resonance profile,
with two peaks becoming evident.  This structure is a signature of quantum effects which are not able to be fully accounted for by the semiclassical approximation to stochastic electrodynamics. The stochastic electrodynamics results do capture
the peak at higher $\bar{\Delta}$ reasonably well, however, the approximation to factorizing two-body correlation functions cannot reproduce the
structure of the local minima and the peak at lower $\bar{\Delta}$.  As the
intensity increases still further, the superradiant peak broadens, the structure
becomes less significant  and the stochastic results once again give good
quantitative agreement.

\begin{figure}
\includegraphics[width=0.48\columnwidth]{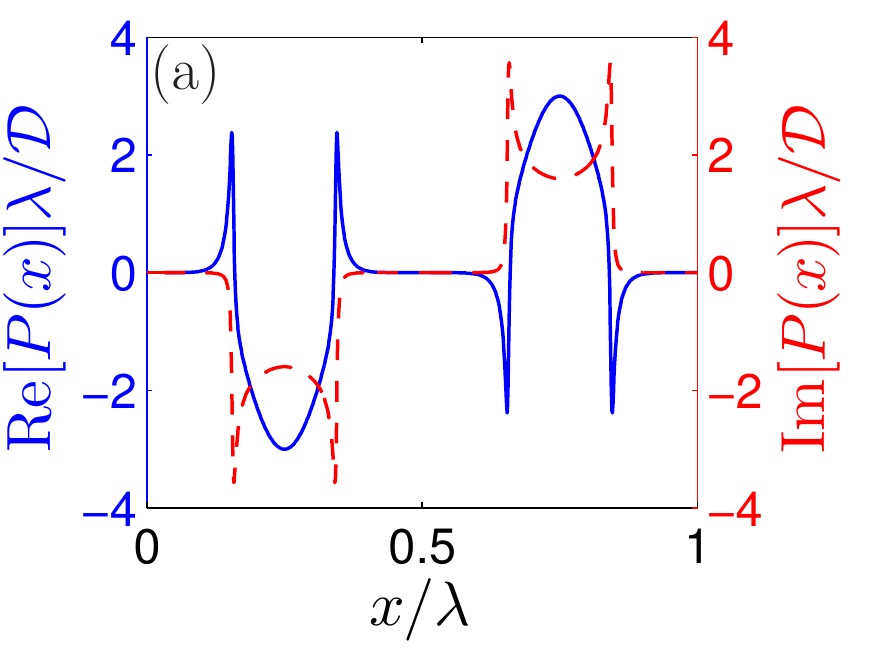}
\includegraphics[width=0.48\columnwidth]{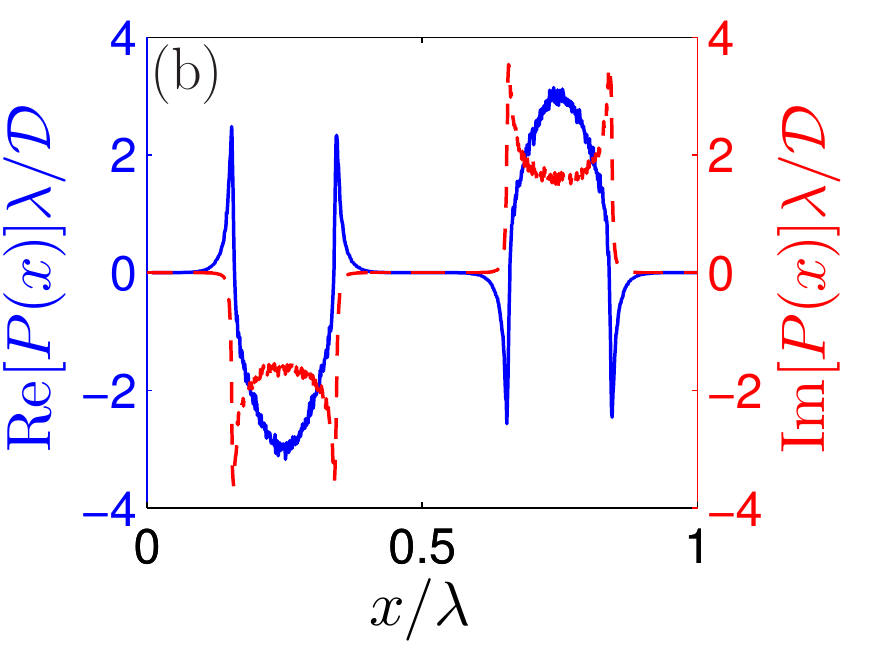}\\
\includegraphics[width=0.48\columnwidth]{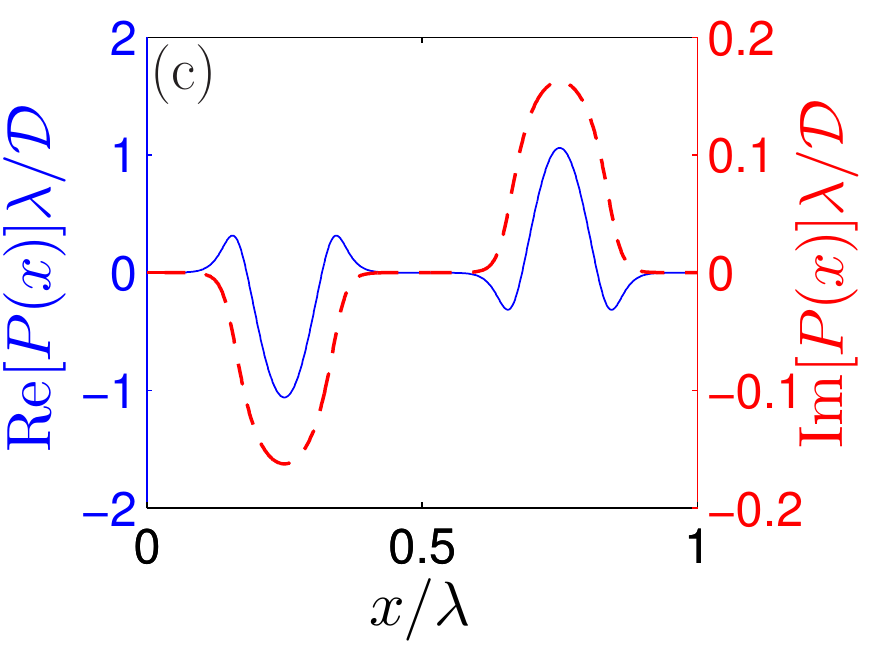}
\includegraphics[width=0.48\columnwidth]{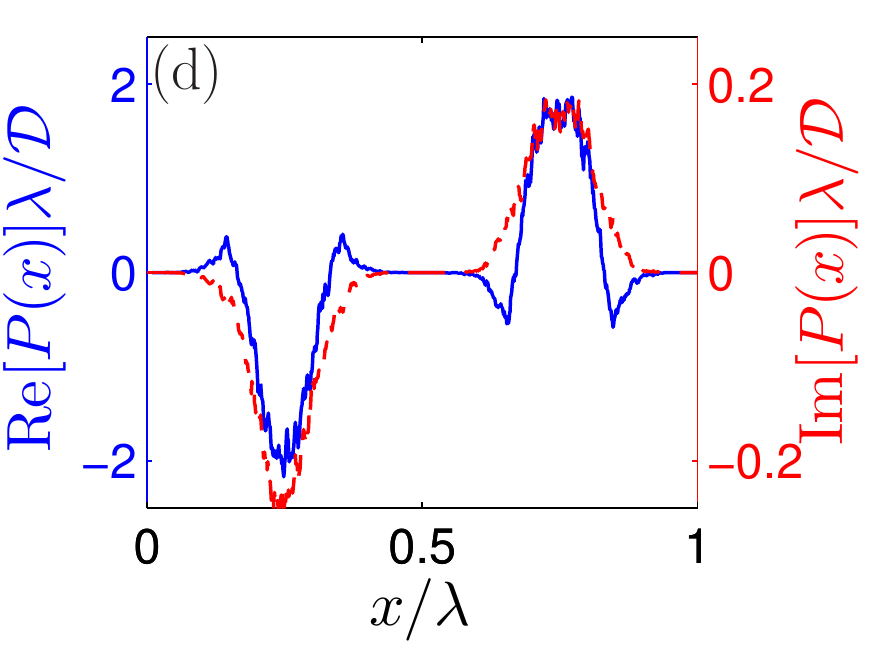}
\caption{Spatial profile of polarization density with the incident light at the resonance of the superradiant
eigenmode for two atoms.  The real (blue, solid) and imaginary (red, dashed)
components of $P(x)$ are shown calculated using  full quantum field-theoretical
treatment (left column) and the semiclassical approximation to stochastic electrodynamics (right column) for a frequency
of $\bar{\Delta}_0 = 4.74\omega_R$. In the limit of low light intensity (top row)
the two methods yield the same result (apart from sampling noise).  The cusp-like profile is caused by the
spatial cavity Lamb shift causing only a narrow spatial region to shift to
resonance at any given frequency.  As the resonance is scanned in frequency from
low to high, the position of the cusps moves towards the center of the atomic
density where the cavity Lamb shift is greater.  The second row shows the case
at higher intensities ($\eta = 0.1\kappa$), the effects of saturation broaden
the cusps in the profile found in the full treatment.  Since the semiclassical approximation fails to describe the quantum features of the resonance profile in this frequency range (see
Fig.~\ref{fig:ExactvsStoc_superradiant}, it also does not give quantitative results
for the spatial profile of $P(x)$.  It does, however, give a reasonable
qualitative agreement, with the right hand plot showing a similar profile to the
full treatment but at a slightly shifted frequency ($\bar{\Delta}_0=3.16\omega_R$).
\label{fig:N2Pspatialstructure}}
\end{figure}

An advantage of the semiclassical approximation to stochastic electrodynamics is the ability to include spatial correlation effects
in a tractable manner even in larger systems.  Spatial variation in atomic density, cavity coupling
strength and detuning can lead to nontrivial effects in the polarization
response of the atoms.  For example, since the cavity Lamb shift [${\cal
D}^2|g(x)|^2\Delta_c/(\kappa^2+\Delta_c^2)$] is a spatially varying quantity, for
a given pump frequency only certain values of $x$ are exactly in resonance with
the local superradiant mode.  The polarization response $P(x)$ of the atoms
therefore exhibits a nontrivial spatial structure, which changes as a function
of detuning from the superradiant mode frequency.
Figure~\ref{fig:N2Pspatialstructure} shows the polarization density at a single
frequency close to the superradiant resonance frequency, with qualitative agreement between the two approaches
even at significant saturation.

The spatial nature of the cavity coupling strength also affects the frequency of
the eigenmode.  One might expect to estimate the mode position by a simplified two-atom model in which spatial variations were averaged.  When $\Delta_c$ is
sufficiently large that $\bar{\kappa}$ may be treated as approximately constant,
this estimate of the superradiant resonance frequency is found by integrating
over Eq.~\eqref{eq:cavitydiscretelowlight} and solving the resulting eigenvalue
problem, resulting in a resonant frequency of $2{\cal D}^2
\mbox{Im}\left[g^2_{\rm av}/\bar{\kappa}\right]$, where $g^2_{\rm av} = \int
g^2(x)\rhoC_1(x) dx/2$ is the average cavity coupling strength experienced by an
atom.  In fact, this estimate is not particularly good; a much better estimate
arises from setting $g^2_{\rm av} = g_0^2$, i.e., setting the coupling parameter
to the maximum experienced by an atom, rather than the average.

\subsection{Subradiant mode}

\begin{figure}
\includegraphics[width=0.49\columnwidth]{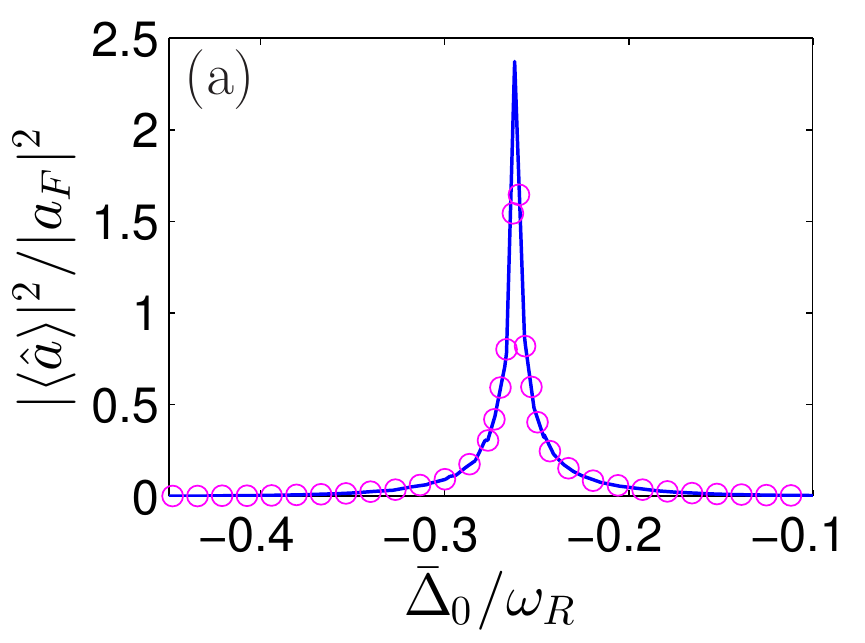}
\includegraphics[width=0.49\columnwidth]{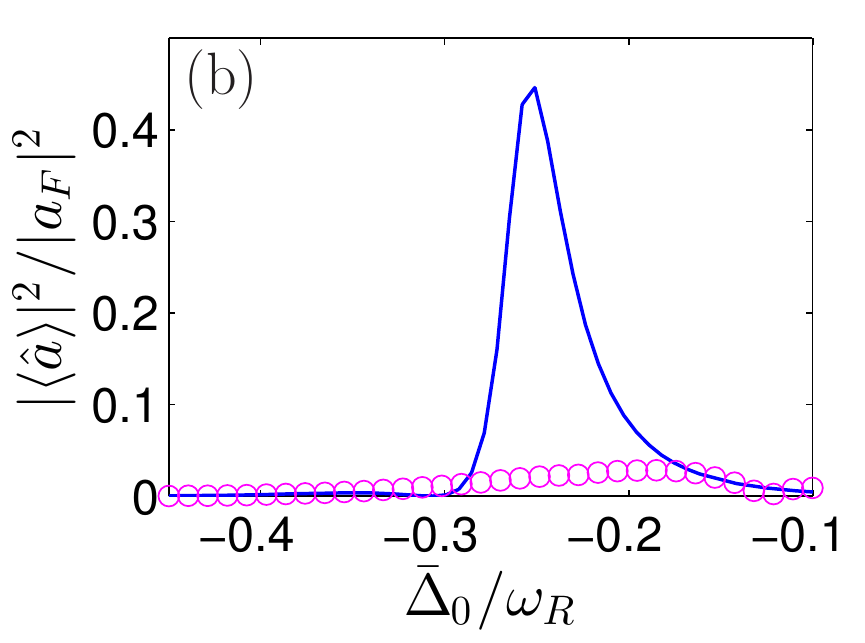}\\
\includegraphics[width=0.49\columnwidth]{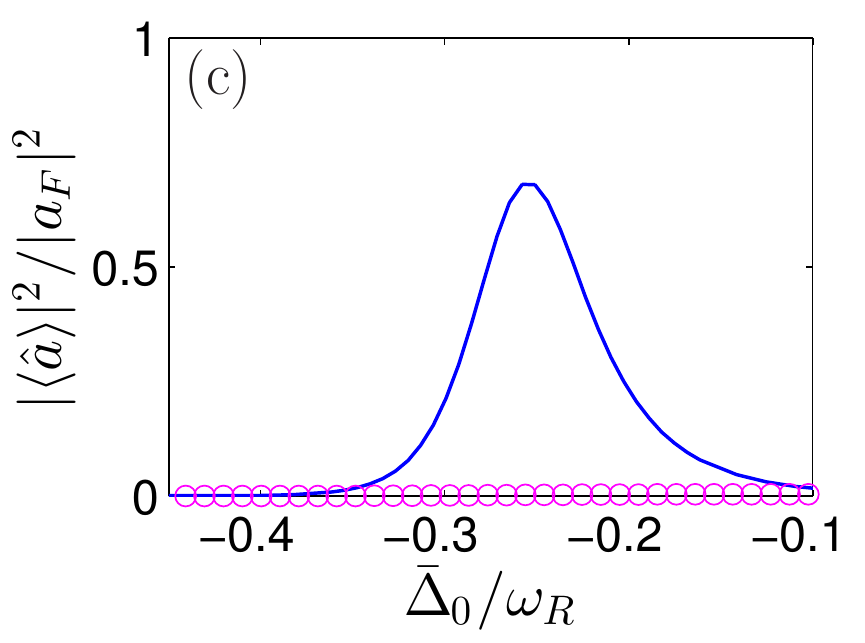}
\includegraphics[width=0.49\columnwidth]{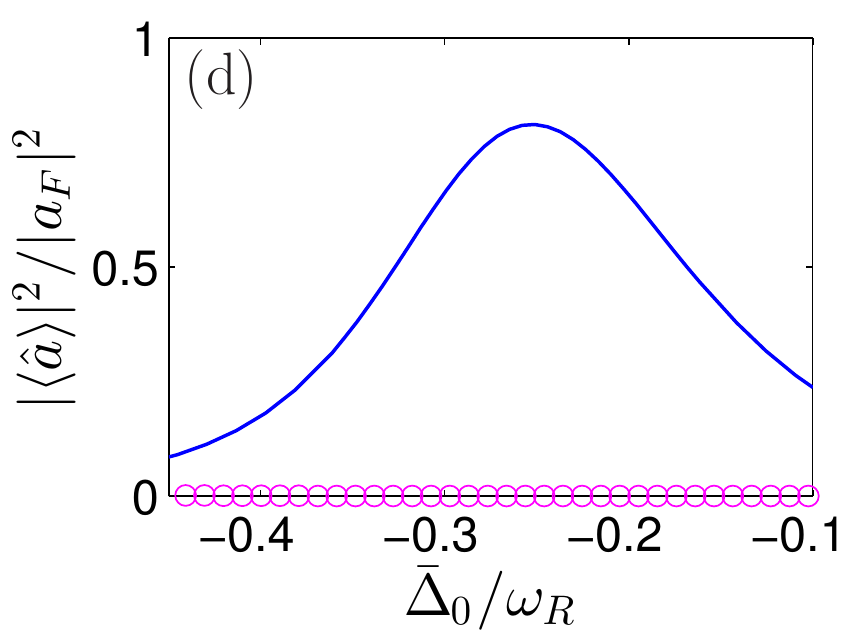}\\
\includegraphics[width=0.49\columnwidth]{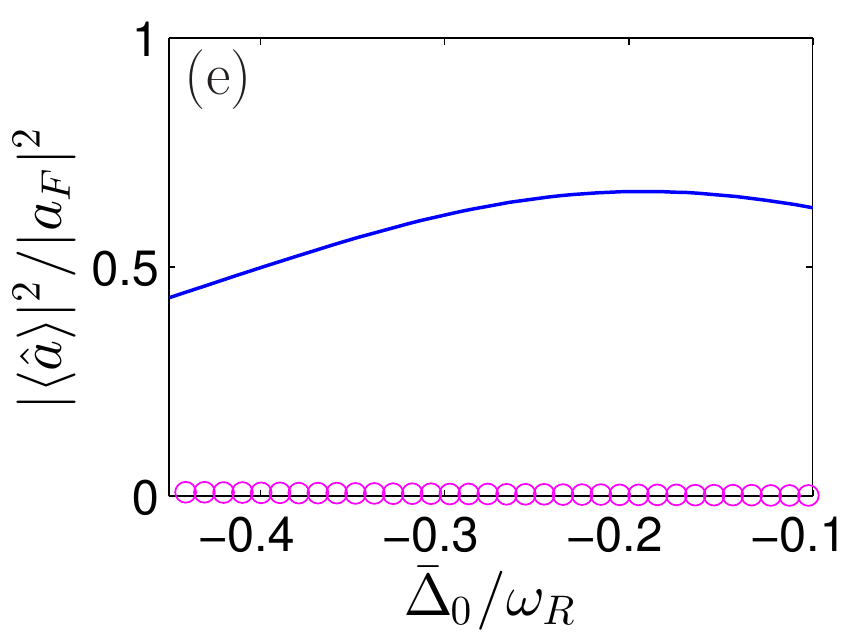}
\includegraphics[width=0.49\columnwidth]{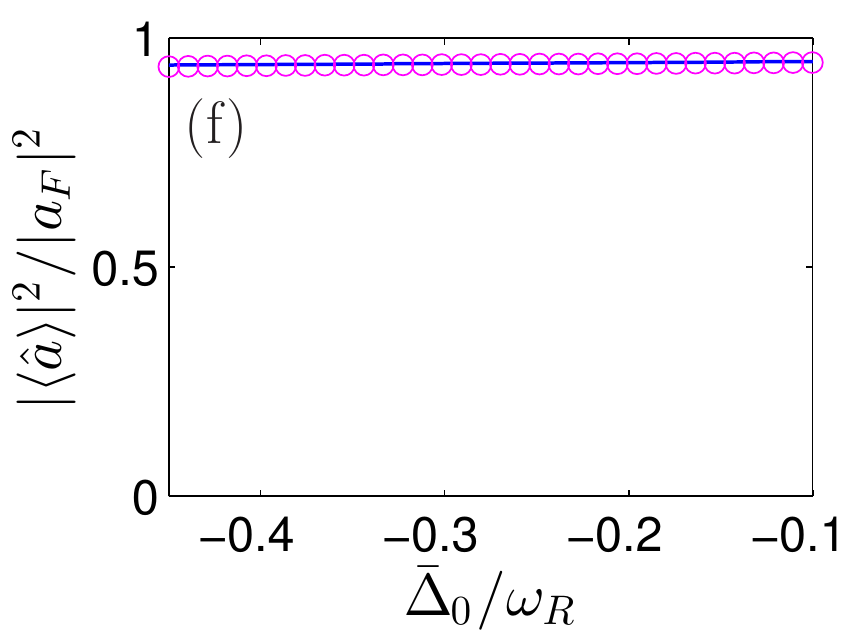}
\caption{The spectrum of the steady-state light intensity inside the cavity for
two atoms showing the resonance due to the subradiant mode.  A shift of
$0.5\omega_R$ has been added to the detuning of the atom in the right hand side
well in order to enable the subradiant mode to couple to the cavity mode. 
Results are shown for full quantum field-theoretical results (lines) and
semiclassical approximation to stochastic electrodynamics (circles) for  the low
intensity limit (a), and the broadening of the resonance as pump intensity
increases outside of the low intensity limit (b)-(f), corresponding to pump
strengths $\eta$ of $0.05\kappa$, $0.1\kappa$, $0.2\kappa$, $0.4\kappa$ and
$5\kappa$ respectively.  The semiclassical stochastic treatment agrees with the
full quantum field-theoretical results in the limit of low intensities, however once
saturation becomes significant the stochastic approximation is unable to capture the
subradiant profile, and already at $0.05\kappa$ the resonance has diminished and
broadened such that it is barely resolvable. At very high intensities, once no subradiant mode is
discernible, the semiclassical approximation again become
accurate.\label{fig:N2Exact_subradiant}}
\end{figure}

Excluding the internal structure of the mode spectral response, the semiclassical approximation to stochastic electrodynamics gives reasonable qualitative agreement for the position and
width of the superradiant mode.  In contrast, however, once saturation becomes
important the semiclassical approximation fails to describe the subradiant mode well.
Figure~\ref{fig:N2Exact_subradiant} shows the spectrum of the steady-state optical cavity response for the two-atom system
for different values of the pump strength. Here, in order to
provide some coupling between the subradiant collective modes and the cavity mode (see
Sec.~\ref{sec:eigenmodes}), we have added a
differential atomic level shift between
the two sites
\begin{equation}
\bar{\Delta}(x) =\bar{\Delta}+0.5\omega_R\theta(x)
\end{equation}
where $\theta(x)$ is the Heaviside function and the center of the two wells is at
$x=0$.
By construction, the semiclassical stochastic approach coincides exactly (within the
sampling error) with the full
field-theoretical solution in the low intensity limit, displaying a resonance peak
over an order of magnitude narrower than that corresponding to the superradiant
mode.  In contrast to the superradiant mode, increasing the intensity does not
lead to any additional internal structure of the resonance profile in the results from the full quantum treatment. However, the subradiant mode does broaden substantially as the
saturation increases.  The semiclassical approximation clearly overestimates the
broadening and resonance shift once saturation is included. At very high intensities, the subradiant mode becomes
indistinguishable from the background created by the broadened superradiant mode
and in this limit the semiclassical approximation again provides good quantitative
agreement.

\subsection{Detuned from collective resonances}

\begin{figure}
\includegraphics[width=0.49\columnwidth]{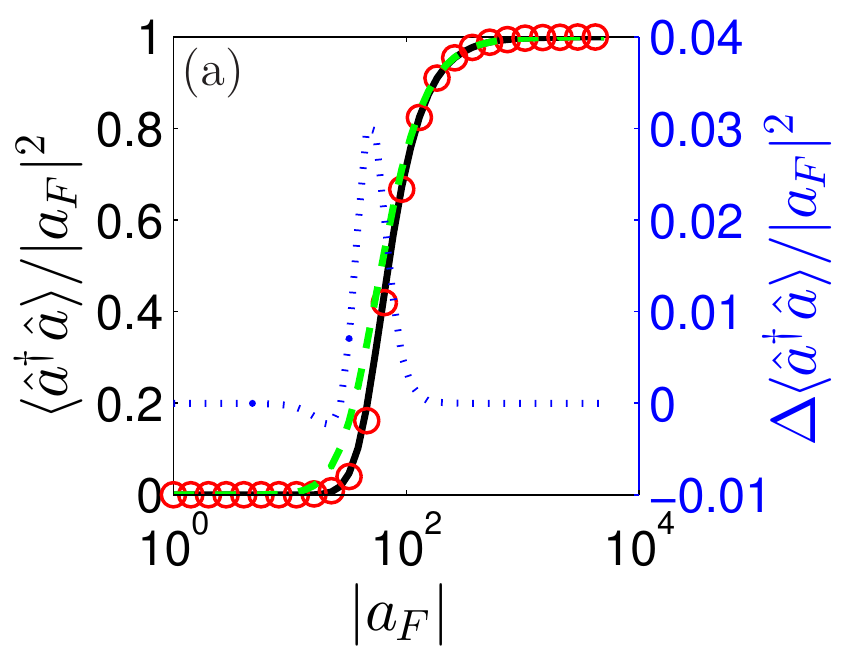}
\includegraphics[width=0.49\columnwidth]{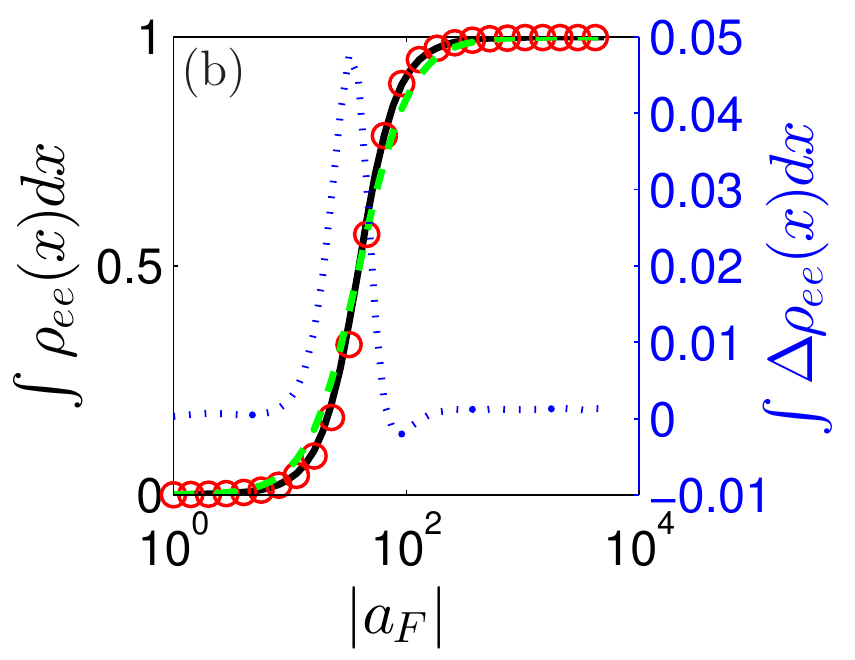}
\caption{Comparison for two atoms in a MI state in a
cavity; full quantum field-theoretical results obtained by solving the hierarchy
of equations of motion for correlation functions compared to the results of the
stochastic electrodynamics simulations. (a) Steady-state intracavity light
intensity $\av{\copa\dopa}$ as a function of driving strength, the coherent part
of the intensity is plotted for the results of the full hierarchy (solid line),
stochastic results with $10^5$ realizations (red, circles), and results from
simple model of a single atom with coupling strength $g_0
\sqrt{N_{\mathrm{eff}}}$ (green, dashed line).  To highlight the discrepancy
between the solution of the full hierarchy and the stochastic simulations with
$10^5$ realizations, the difference between them $\Delta\av{\copa\dopa}$ is
plotted on the right-hand axis (blue dotted line)   (b) Steady-state total
excited state population $\int\rho_{ee}(x)\mathrm{d}x$ for the two atom system
[lines as in (a)], where $\rho_{ee}(x) =
\av{\cpsi_e(x)\dpsi_e(x)}$. The
difference in excited state population between the full hierarchy solutions and
the stochastic simulations (right-hand axis). The system was pumped along the
cavity axis, on resonance with the cavity mode, and the parameters used were
$g_0\Dc/\kappa = 42$, $\bar{\Delta}/\kappa = -4.2$.
\label{fig:Cavity2atom}}
\end{figure}

Detuned from any particular collective resonance, the semiclassical approximation to stochastic electrodynamics
works comparatively well at all studied intensities.  Figure~\ref{fig:Cavity2atom} shows
a comparison with the full hierarchy of equations of motion over a range of
saturation strengths when the incident light is detuned from any of the collective eigenmode resonances.  Agreement is excellent at high values of saturation, although at
intermediate intensities the semiclassical approximation does lead to a small
discrepancy with the full treatment. It is instructive to compare these with other
approximate treatments. For $N$ atoms in a cavity, each of which interact with the
cavity mode with identical coupling strengths $g_0$, the system reduces to the
Tavis-Cummings model~\cite{Tavis1968a}, and in the limit of low light intensity
this simplifies to give the equivalent response to that of a single atom with
coupling strength $g_0\sqrt{N}$.  Motivated by this, a common simple approximate
treatment of spatial variations of coupling strengths $g(x)$ is to solve a
similar model of a single atom with coupling strength $g_0\sqrt{N_{\mathrm{eff}}}$, where $g_0^2
N_{\mathrm{eff}} = \int g^2(x) \rhoC_1(x)\mathrm{d}x$.  The results of this
approximation are also shown in Fig.~\ref{fig:Cavity2atom}, and while they tend
to the correct limit for low light intensity, they are in general less accurate than
the results of the semiclassical electrodynamics simulations.

\subsection{Comparison between correlation functions}

In the semiclassical
model for each stochastic realization we solve the equations of motion for
single-particle variables for internal atomic states. Ensemble-averaging over
many realizations of atomic positions establishes spatial correlations between the atoms.
However, not all the correlations that result from the full quantum dynamics are
incorporated. This can be identified by comparing in more detail the equations for
internal atomic states for each stochastic realization in the semiclassical approximation
with the full quantum dynamics of the internal states for fixed atomic positions.

In
Eqs.~\eqref{eq:eqnmotionP2cavity2atoms} and~\eqref{eq:eqnmotionP+P-2atoms} we give
examples of the full quantum equations of motion for two of the two-body correlation functions
$P_2(x;x')$ and $\av{:\Pop^+(x)\Pop^-(x'):}$, where $:\cdot:$ represents normal
ordering.  The semiclassical factorization approximation together with the stochastic
sampling technique synthesizes two-body correlation functions from the products of
one-body quantities evaluated for each stochastic realization.  For a single realisation
of atoms at the discrete positions $X_j$ and $X_l$, one can obtain the implied
semiclassical equations of motion for the products $\rho_{gg}^{(j)}\rho_{ge}^{(l)}$ and
$\rho_{ge}^{(j)}\rho_{eg}^{(l)}$ [using Eqs.~\eqref{eq:cavitystochastictwolevel}].  On
the other hand, full quantum equations of motion for the two-body quantities $\rho_{gg;ge}^{(j,l)}$ and
$\rho_{ge;eg}^{(j,l)}$ can be derived from Eqs.~\eqref{eq:eqnmotionP2cavity2atoms} and Eqs.~\eqref{eq:eqnmotionP+P-2atoms},
respectively, evaluated at the same discrete positions, by substituting
\begin{align}
\langle\hat\psi^\dagger_g(x) \hat\psi^\dagger_g(x')\hat\psi_e(x)&
\hat\psi_g(x)\rangle_{\{X_j,X_l\}}\nonumber
\\ &=\rho_{gg;ge}^{(j,l)}
\delta(x-X_j)\delta(x'-X_l)\nonumber \\&+\rho_{gg;ge}^{(l,j)}
\delta(x-X_l)\delta(x'-X_j)\,,
\end{align} 
and
\begin{align}
\langle\hat\psi^\dagger_g(x) \hat\psi^\dagger_e(x')\hat\psi_g(x)&
\hat\psi_e(x)\rangle_{\{X_j,X_l\}}\nonumber
\\ &=\rho_{ge;eg}^{(j,l)}
\delta(x-X_j)\delta(x'-X_l)\nonumber \\& +\rho_{ge;eg}^{(l,j)}
\delta(x-X_l)\delta(x'-X_j)\,,
\end{align} 
and similarly for all other two-body correlation functions.  Subsequently factorizing all
two-body terms in the resultant equations of motion [in the manner of
Eq.~\eqref{eq:cavityfactorization}] leads to
terms consisting of products of one body quantities. Comparing these equations with those
obtained by making the factorization approximation at the outset gives some insight into
the nature of the semiclassical approximation.

For both of these two-body correlation functions, the semiclassical equations of motion
reproduce the diagonal and driving terms (those proportional to $a_F$ or $h(x)$) from the
full quantum treatment.  In the case of $d\rho_{ge;eg}^{(j,l)}/dt$ the off-diagonal term
proportional to $\rho_{ee}^{(j)}\rho_{ge}^{(l)}$ is included, while the terms missing
from the semiclassical factorized version are identical to those included but with a swap
of coordinates $X_j \leftrightarrow X_l$. A crucial aspect of the stochastic treatment is
the averaging over many such stochastic realizations. In the low intensity limit it can
be shown that, although not present in the factorized version for a single realization, 
the swapped coordinate term proportional to $P_2(x';x)$ in Eq.~\eqref{eq:eqnmotionP2cavity2atoms} is included by means of averaging
over the stochastic samples \cite{Lee16}.

In contrast, the semiclassical approximation to the equation of motion for
$\rho_{ge}^{(j)}\rho_{eg}^{(l)}$ does not reproduce any of the off-diagonal couplings
present in the full result for $\rho_{ge;eg}^{(j,l)}$.  Since in general in the full
quantum picture all two-body correlation functions are coupled, the missing terms mean
the semiclassical approximation cannot reproduce the quantum correlations, even
when averaged over multiple realizations.  However, in the low intensity limit the only
two-body correlation functions which are important for the cavity optical response are
$P_2(x;x')$ and $P_2(x';x)$, and hence the same factorization approach is able to fully
reproduce the correlations in this limit.

\begin{figure}
\includegraphics[width=0.49\columnwidth]{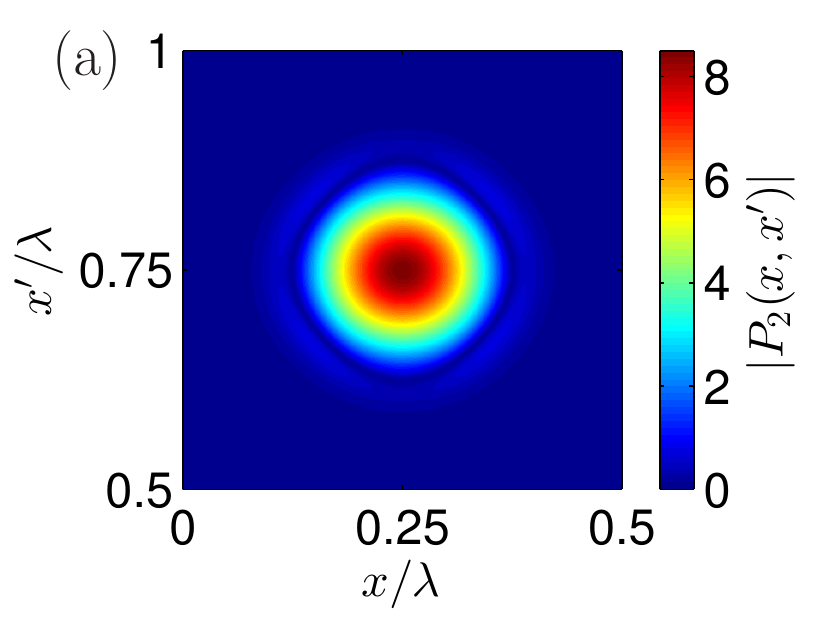}
\includegraphics[width=0.49\columnwidth]{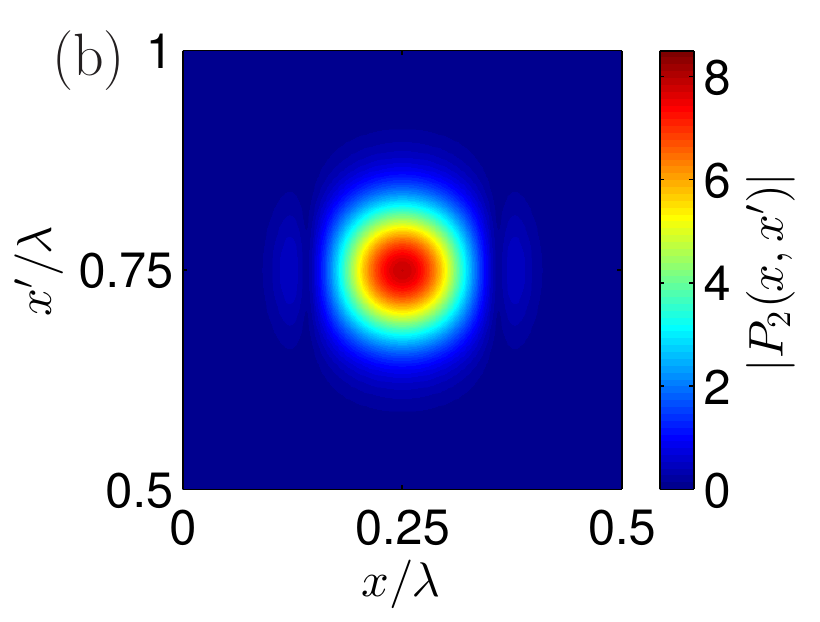}\\
\includegraphics[width=0.49\columnwidth]{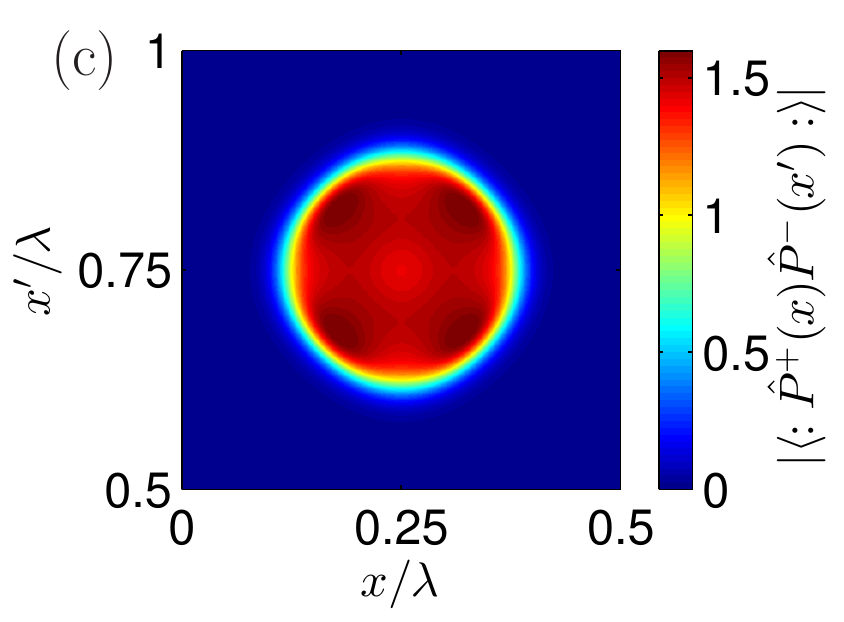}
\includegraphics[width=0.49\columnwidth]{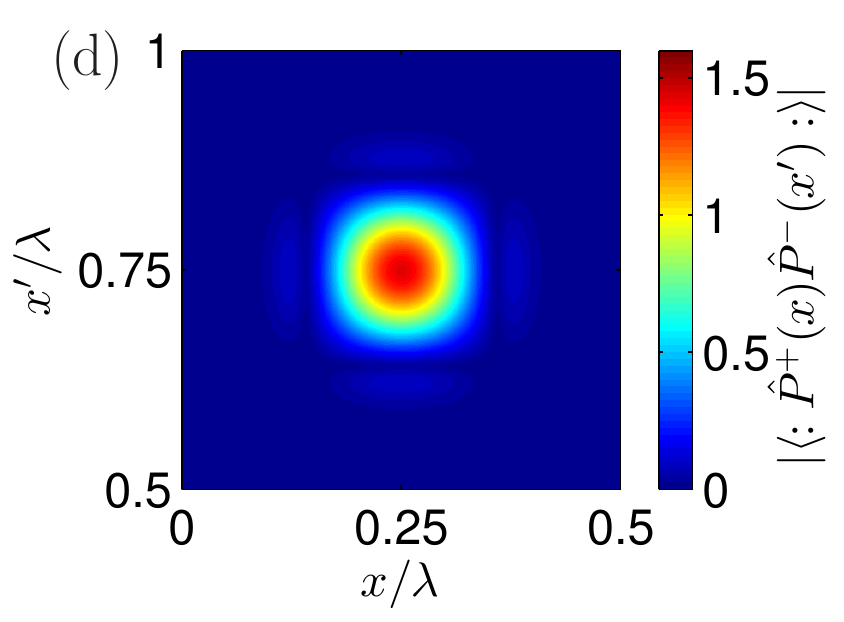}\\
\includegraphics[width=0.49\columnwidth]{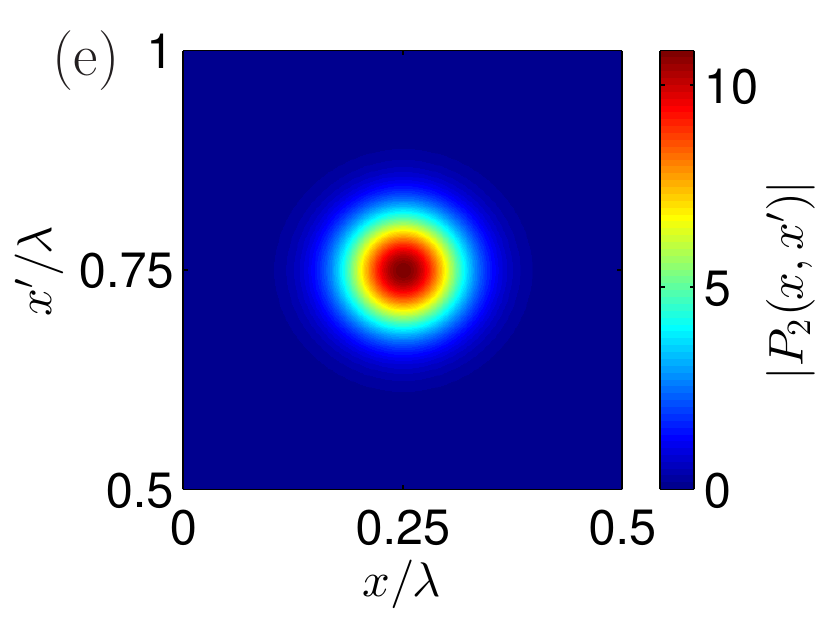}
\includegraphics[width=0.49\columnwidth]{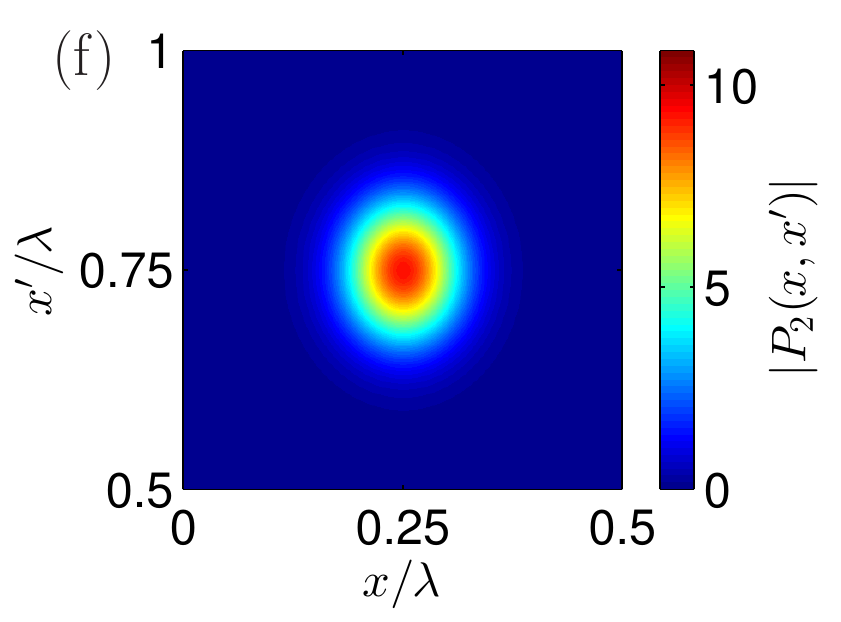}\\
\includegraphics[width=0.49\columnwidth]{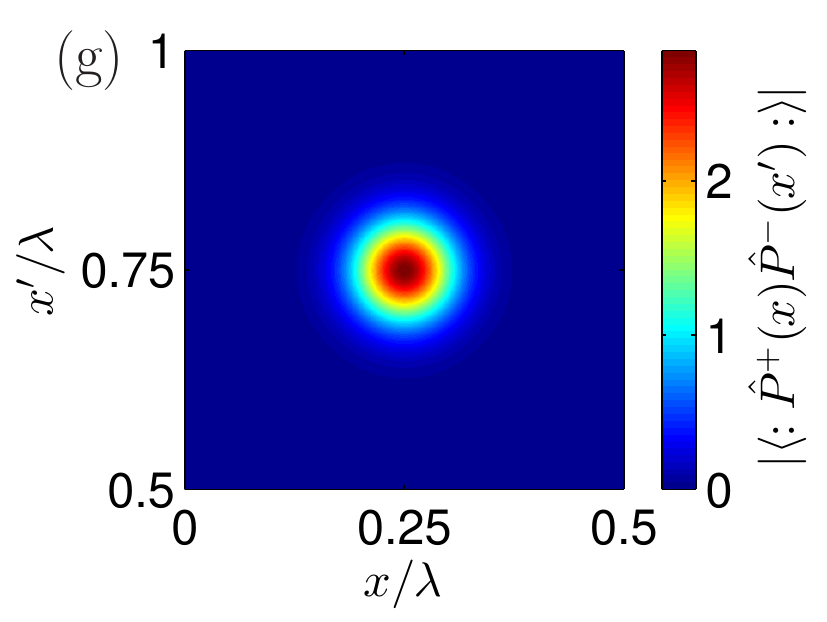}
\includegraphics[width=0.49\columnwidth]{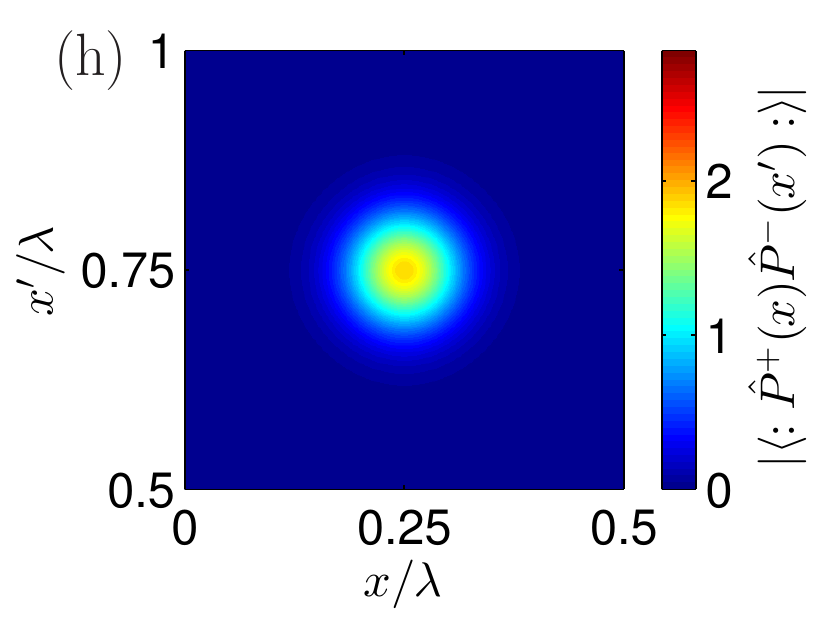}
\caption{Magnitude of two-body correlation functions at two different frequencies within
the range of the superradiant mode resonance.  We show the magnitude of two of the two-body
correlation functions, $P_2(x;x')$ and
$\av{:\Pop^+(x)\Pop^-(x'):}$, calculated from the
results of the full treatment of all two-body correlation functions.  The left
column shows the full quantum result, while the right hand column shows the semiclassical approximation
to the correlation functions in the stochastic electrodynamics.  The upper two rows show the
behavior at a frequency corresponding to the left hand peak of
Fig.~\ref{fig:ExactvsStoc_superradiant}, and it can be seen that the semiclassical
approximation is less accurate for this frequency.  In contrast, the semiclassical
approximation works significantly better in the lower two rows, obtained at a
frequency corresponding to
the right hand peak of Fig.~\ref{fig:ExactvsStoc_superradiant}.  These results
were obtained with $\eta = 0.1\kappa$.  Note that due to the symmetry of the
MI system we show only one quarter of the full domain.
\label{fig:N2super_corrfunc}}
\end{figure}

Figure~\ref{fig:N2super_corrfunc} compares two different two-body correlation
functions obtained from the full treatment of the hierarchy of correlation
functions with factorized semiclassical approximations at two different frequencies near
resonant with the superradiant eigenmode and within the range spanned by
Fig~\ref{fig:ExactvsStoc_superradiant}.  It can be seen that when the semiclassical approximation to stochastic
electrodynamics is least accurate, the full two-body correlation functions exhibit
structure that cannot be reproduced by any factorization approximation.  In
contrast, where the factorized and full correlation functions show similar
qualitative features the semiclassical approximation agrees rather well with
the full treatment for the optical response of this collective mode.

In summary, while the semiclassical approximation to stochastic electrodynamics has no effect in the limit of low light intensity
and the two methods are the same (beyond the sampling noise), at intermediate intensities when saturation effects become important the
semiclassical approximation only gives qualitative agreement for the excitation of the superradiant eigenmode and is unable to
describe the excitation of the subradiant eigenmode. In the limit of very high intensities,
quantitative agreement is restored although the collective features have become
substantially power-broadened in this limit.  Detuned from the collective modes,
the semiclassical approximation works well at all intensities.

The semiclassical approximation to stochastic simulations is designed to capture spatially-dependent correlation effects induced by the scattered light.
In cavities such effects are weaker than in free space, owing to the unattenuated long-range dipole-dipole interactions between the atoms. Furthermore,
light confinement and directed light emission enhance quantum effects in cavities compared with quantum fluctuations in free space. Consequently, finding differences
in the cavity response of a small two-atom system between the semiclassical approximation and the exact quantum result is not entirely surprising, but such deviations
are likely to become smaller in larger atomic ensembles and in multimode cavities.

\section{Diagnostics of atomic quantum phase}
\label{sec:manybody}

An advantage of the stochastic electrodynamics simulations is that different
atom distributions and statistics are incorporated in the joint
probability distribution from which the stochastic realizations of discrete
atomic positions are sampled.  In many situations an independent atom sampling can be employed;
for example, for an ideal BEC  or for uncorrelated (in the absence of light) classical atoms confined in an optical lattice potential the
positions are sampled from a distribution proportional to the total linear
density $\rhoC_1(x)  = \sum_i n_i|\phi_i(x)|^2 dx$ where $\phi_i(x)$ is the
wavefunction for site $i$ with site population $n_i$.  In
contrast, we described how to simulate a Mott Insulator (MI) state in
Sec.~\ref{sec:eigenmodes}.

The atom statistics can affect the cavity system response when site-to-site spatial variation
in the Wannier functions, atomic detunings or cavity coupling strength are
present.  The optical response can therefore be used in principle as a
diagnostic tool to query the quantum phase of the atoms.  As another illustration of the stochastic electrodynamics simulations we discuss two
examples below where the MI and BEC states differ: firstly, the distribution of
subradiant modes, and secondly the case of a cavity transmission spectrum when
some sites are masked from the cavity, following the model of Ref.~\cite{Mekhov2007a}.
Off-resonance spontaneous scattering in optical lattices has actively studied as a diagnostic tool for quantum and thermal states of atoms; see
for instance Refs.~\cite{Javanainen03,Ruostekoski09,Rist10,Trippenbach,JamesBurnett10,JamesBurnett11,Weiping11,Sandner,CordobesAquilar,Elliott15}.

\subsection{Subradiant modes for a BEC}

In Fig.~\ref{fig:N8eigenmodes} we presented the subradiant modes which appear
when $\bar{\Delta}(x)$ was not spatially constant but had a weak constant
gradient, for a system of $8$ atoms in a MI state of 1 atom per lattice site.
Seven subradiant and one superradiant modes are clearly (in this case)
resolvable.  The subradiant modes are only coupled to the cavity mode when the
detuning $\bar{\Delta}(x)$ varies between lattice sites.  Since different atomic
quantum phases lead to different atom probability distributions, they will in turn
lead to different sampled detunings in the possible stochastic realizations of
atom positions.  The nature of the subradiant modes might therefore be expected to
vary with the atomic quantum phase.

\begin{figure}
\includegraphics[width=0.49\columnwidth]{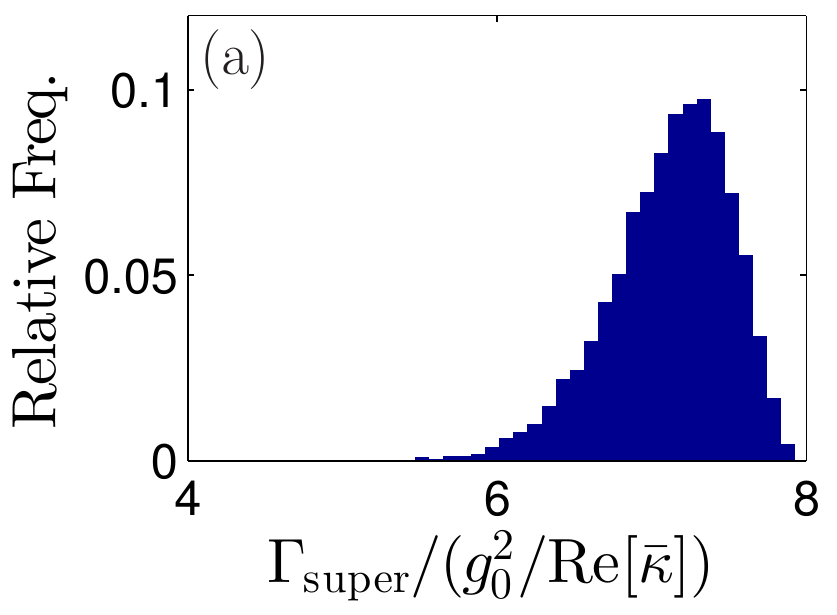}
\includegraphics[width=0.49\columnwidth]{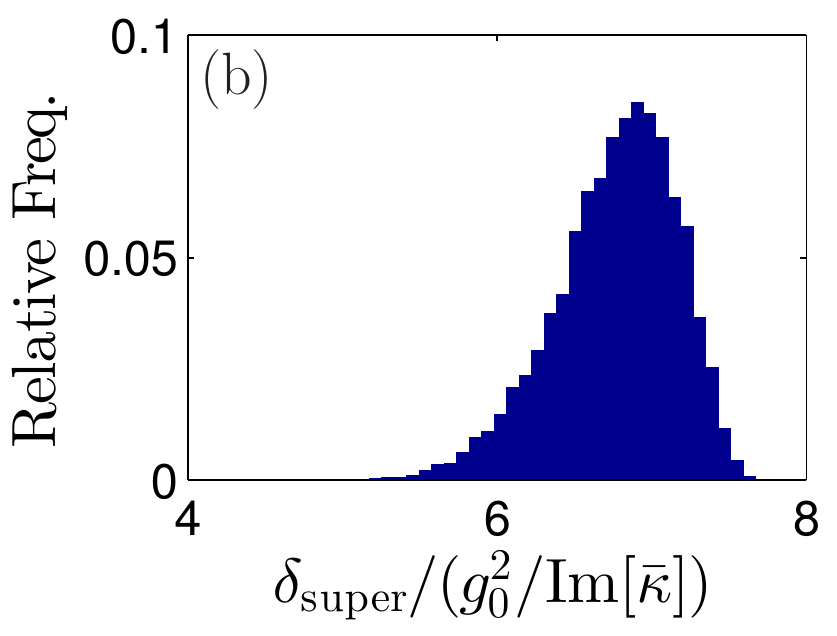}\\
\includegraphics[width=0.49\columnwidth]{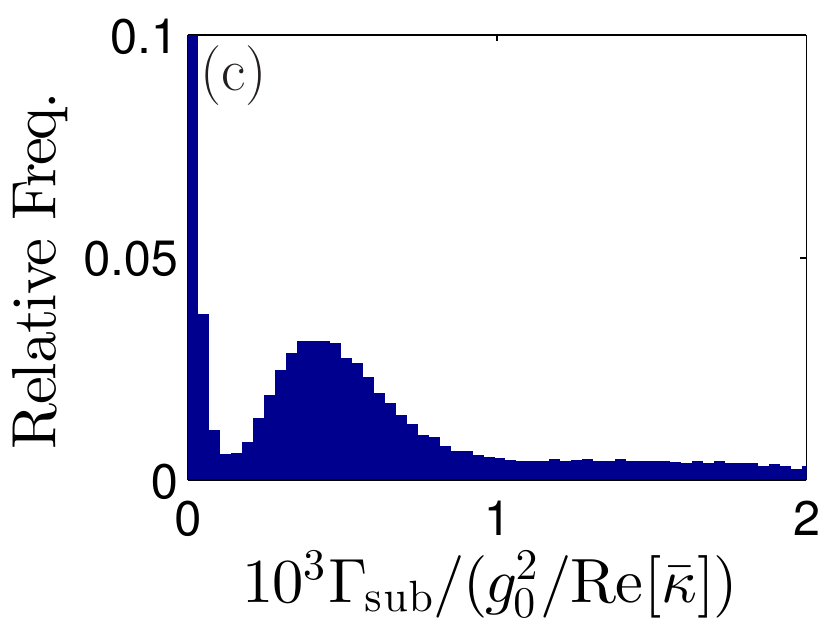}
\includegraphics[width=0.49\columnwidth]{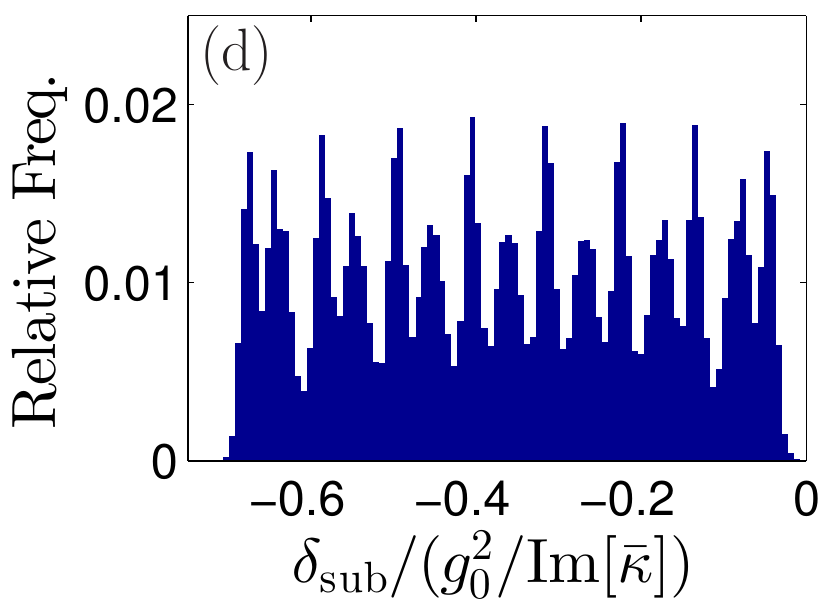}
\caption{Distribution of eigenmode decay rates $\Gamma$ and
frequencies $\delta$, for an ensemble of individual realizations of stochastic
atomic positions for 8 atoms in a superfluid state in an optical lattice.  The
superradiant mode is shown in (a) and (b) while the $7$ subradiant modes are
illustrated in (c) and (d), since it is now difficult to sort the individual
subradiant modes only the overall distribution is shown.
System parameters are identical to those for Fig.~\ref{fig:N8eigenmodes}.
\label{fig:SFsubradiant}}
\end{figure}

Indeed, if we instead consider the same system as Fig.~\ref{fig:N8eigenmodes}
but in a superfluid BEC phase, the site number fluctuations significantly alter
both the lifetimes and resonance frequencies of the subradiant modes.  Modes are
no longer easily separable in frequency, and the distribution of lifetimes
acquires a two peak structure, as shown in Fig.~\ref{fig:SFsubradiant}.  In
contrast, the superradiant collective mode is not significantly affected by the
atomic quantum phase, since it is not greatly affected by any small changes in
$\bar{\Delta}(x)$. For simplicity, we have assumed that the lattice potential is
unaffected by the spatially dependent $\bar{\Delta}(x)$, so that the Wannier
functions for each lattice site are identical.

\subsection{Number fluctuations in the cavity transmission spectrum}

A clear example of the importance of the atomic quantum phase was presented by
Mekhov, Maschler and Ritsch~\cite{Mekhov2007a} when they considered the
transmission spectra of a cavity containing an optical lattice commensurate with
the cavity mode, but such that only $n_A \leq n_L$ of the total $n_L$ sites
occupied by atoms interact with the cavity mode (Ref.~\cite{Mekhov2007a} proposes
to achieve this by tilting the axis of the 1D optical lattice with respect to the
cavity axis).

In contrast to most of the results in this paper, we concentrate in this section
on the dressed mode of the cavity-atom system which is dominated by the bare cavity mode
(illustrated by the red dashed line in Fig.~\ref{fig:DressedStatesScan}).  We work in the
regime where the bare atomic transition is detuned far from resonance with the
driving axial cavity pump, but where the
cavity mode itself is near resonant with the pump  ($|\bar{\Delta}|\gg|\Delta_c|$). In the
absence of atoms, the empty cavity exhibits a resonance peak in the transmitted
photon intensity centered at $\Delta_c=0$ with a width dictated by the cavity
loss rate $\kappa$.  The presence of far-detuned atoms in the cavity shifts
the position of this resonance, proportional to the number of atoms in the
cavity (with reference to Fig.~\ref{fig:DressedStatesScan}, the shift is that of
the dressed state of the cavity-atom system from the bare cavity resonance).  This shift, together with
the geometry of the system described above, allows the signature of the atomic
many-body state to be seen in the cavity transmission spectrum.

For a MI state with exactly one atom per site, $n_A$ atoms interact
with the cavity mode to shift the cavity resonance frequency.  For an ideal BEC superfluid
however, site number fluctuations mean that a given experimental realization can
involve any number of atoms between $0$ and $N$, each realization giving rise to
a different resonance shift.  The signature of the superfluid state is therefore
a comb-like transmission spectra, compared to the single peaked response of the
MI state.

\begin{figure*}
\includegraphics[width=0.49\columnwidth]{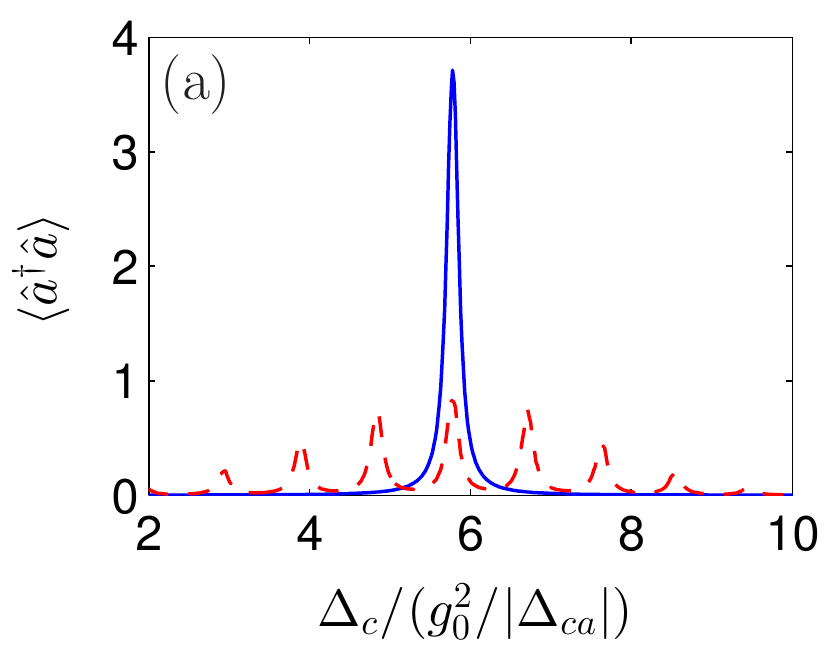}
\includegraphics[width=0.49\columnwidth]{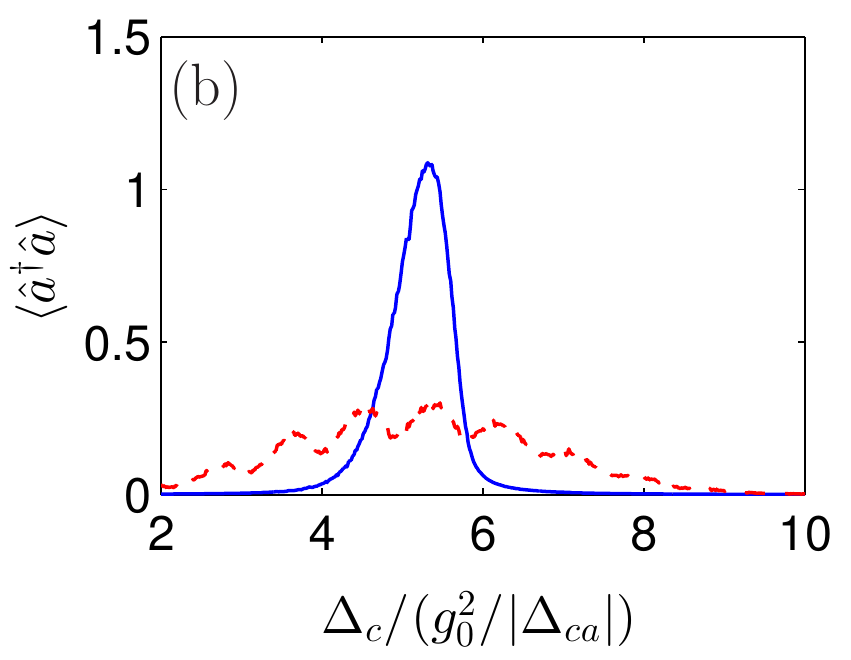}
\includegraphics[width=0.49\columnwidth]{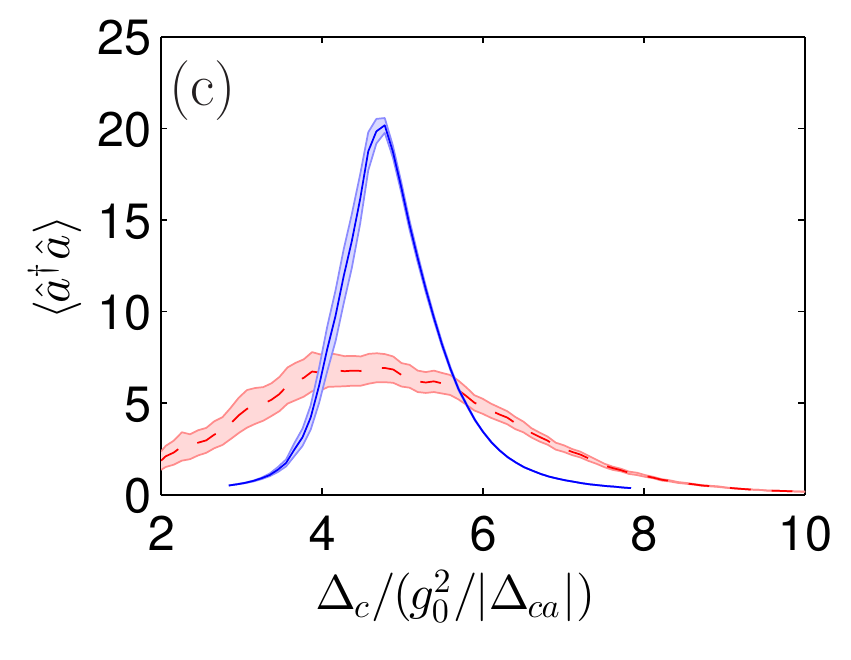}\\
\includegraphics[width=0.49\columnwidth]{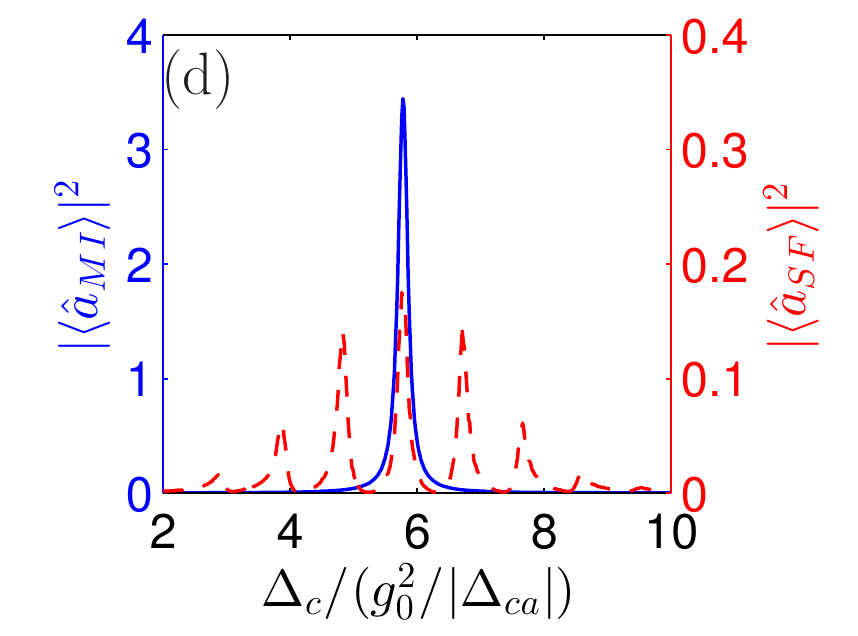}
\includegraphics[width=0.49\columnwidth]{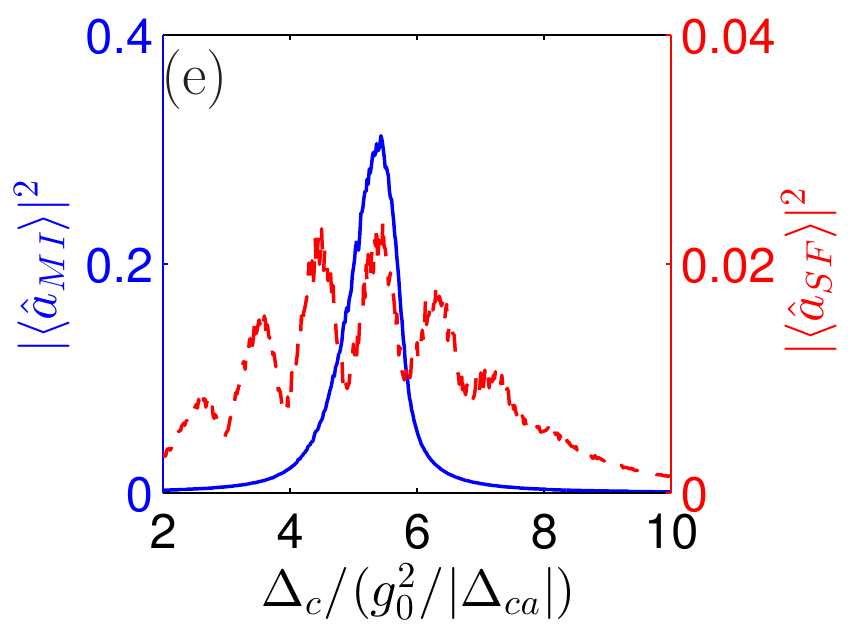}
\includegraphics[width=0.49\columnwidth]{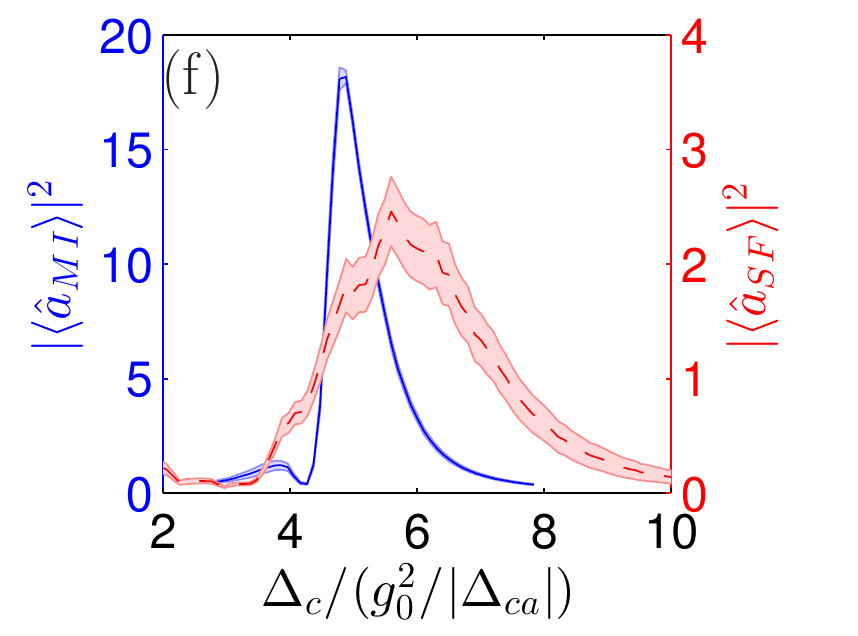}\\
\caption{Effect of number fluctuations on the steady-state cavity transmission
spectrum due to different atomic many-body states.  An optical lattice of $12$
sites and containing $12$ atoms is placed in a cavity, but tilted so that $6$ of
the lattice sites are commensurate with the cavity mode, but the remaining $6$
sites do not interact with the lattice.  The figures then show the transmission
spectrum of the cavity mode resonance, which is shifted from $\Delta_c = 0$ by the
presence of the atoms. (a) In the limit of low intensity, atoms in a MI state show no number fluctuations and give a single peak (blue, solid)
while the number fluctuations from a perfect superfluid state give rise to a comb
pattern (red, dashed).  Here atom densities in each lattice site $\ell_i$ have
relatively narrow distributions $\rhoE{i}{x} \propto
\exp\{-[(x-\ell_i)/(0.02\lambda)]^2\}$ and $\eta/\kappa = 2$.  (b) Increasing  the width of the density
distributions in each site ($\rhoE{i}{x} \propto
\exp\{-[(x-\ell_i)/(0.08\lambda)]^2\}$) causes much of the structure to be washed
out.  The difference between atomic states is now mostly evident in the width of
the spectrum.  (c) At higher intensities, saturation effects further wash out the
earlier structure, here the system is pumped axially with $\eta/\kappa = 20$,
for atom densities as in (b).  The bands represent statistical uncertainties due
to comparatively low number of stochastic realizations, and are not resolvable
in the low intensity results.  (d)-(f) as for (a)-(b), respectively, but showing
the coherently scattered light only $\propto |\< \hat a \>|^2$. Other system parameters for all cases are $g_0/\kappa = 168$, $\kappa =
-0.08g_0^2/\Delta_{ca}$.
\label{fig:numberfluc}}
\end{figure*}

Our stochastic simulations are well suited to tackle this system, and the
contrasting spectra between the two atomic phases are shown in Fig.~\ref{fig:numberfluc}(a)
Here, we have simulated a system of $12$ atoms with in an optical lattice with
narrow Wannier functions and with only half of the sites of the optical lattice
able to interact with the cavity light.
The numerical calculation extends the results of Ref.~\cite{Mekhov2007a} to situations where
the atomic transitions are allowed to be saturated by high pump intensities and the coupling strength
is allowed to vary in space.
In Fig.~\ref{fig:numberfluc}(b) we show results for a lattice where the spatial widths of Wannier
functions are on the order of $\lambda/2$, leading to a range of cavity coupling
strengths $g(x)$ experienced by the atoms.  In Fig.~\ref{fig:numberfluc}(c), we
show the effect of significant
saturation.   For each case, we show both the coherently scattered light transmission $\propto |\< \hat a \>|^2$
and the total intensity  transmission $\propto \< |\hat a |^2\>$  that also includes the incoherently scattered light component.
The two signals for the MI state are very similar, but they differ more in the case of a BEC due to the fluctuating density,
as in the latter case the incoherently scattered light $\propto  \< |\hat a |^2\> - |\< \hat a \>|^2$ has a stronger effect on the optical response.

In comparison to Fig.~\ref{fig:numberfluc}(a), both widening the widths of the Wannier functions and including saturation can be seen to lead to a
loss of resolution in the comb shape for the superfluid response and a broadening
for the MI case.  The larger width of the Wannier function leads to a broadening
of the resonance peaks because the wider atom distribution samples a greater range of
cavity coupling strengths.  Furthermore, since the geometry of the system leads to a
decrease in the total experienced coupling strength, the peak is shifted to
smaller $\Delta_c$.  The distribution of $\sum_j g(X_j)$ sampled is not
symmetric, and this is reflected in the appearance of the resonance peak.
Higher intensities decrease the resolution through power broadening of the
spectra.  Nonetheless, although such considerations may mean the distinctive
comb-like structure is not always resolvable, the differing widths of the
superfluid and MI resonances means that the spectra can still be used as a
diagnostic tool for the quantum phase of the atoms.

\section{Concluding remarks}

We have formulated stochastic electrodynamics for many-atom systems in a cavity.
In this work the approach was implemented in a semiclassical approximation that 
could be extended also to include quantum fluctuations. However, in the limit of
low light intensity and for two-level atoms, the presented simulations are
limited in accuracy only by the sampling error introduced by the finite number
of stochastic samples used. Our formulation of the stochastic simulations has
the advantage that atomic position correlations and spatially dependent
potentials and couplings are readily included.

We have shown that a system of atoms in a cavity can exhibit a collective
optical response with a strongly enhanced superradiant mode, but also with a
number of weakly coupled subradiant modes with very narrow linewidths.  These
subradiant modes can be coupled to the cavity response via spatially dependent
detunings, and can exhibit distinct spatial profiles allowing them to be driven
by tailoring the shape and phase of a transverse optical pump.

This phenomenon also raises a possible mechanism for the storage of light by
many-body atomic excitations in the cavity. Provided that a controllable
spatially-dependent level shift can be introduced for the atoms when the cavity
mode is driven, the modes that are subradiant in the absence of the
spatially-dependent shifts can be directly excited by the driving field. 
The idea is related to the analysis of Ref.~\cite{Facchinetti} where a subradiant
mode was driven in a planar optical lattice in free space by coupling the
different atomic polarization components with the Zeeman level shifts.
After a suitable excitation pulse, the level shifts are turned off, such that the
finite, nonvanishing resonance linewidth of the excited subradiant mode (of the
case with spatially-dependent level shifts) is tuned as close as possible to
zero. The excitation may then become trapped in the subradiant mode in a way
that a decay via the cavity mode may in principle exhibit an extremely long
lifetime. The atoms may still decay via the free space modes perpendicular to
the cavity, but such a decay is sensitive to the spatial arrangement of the
atoms, and the decay can be weak for regularly spaced array of atoms with a
subwavelength lattice constant~\cite{Sutherland1D,Bettles1D}.

The semiclassical stochastic method was compared with the full quantum solution,
revealing quantum features in the optical response. These are particularly
prominent in the excitation of narrow subradiant resonances at high intensities.
We also showed how the atomic quantum phase of the ensemble may be discerned
from the the cavity optical response due to the different atomic position
correlations in differing phases.  The quantum statistics of the atoms inside
the cavity is mapped onto the optical signal, and the statistics is reflected in
different ways in the coherently and incoherently scattered light.

\appendix

\section{Polarization and correlation function equations of motion for
multi-level atoms}
\label{App:AppFullEqns}

Equations~\eqref{eq:cavitydPdt} and~\eqref{eq:cavitydrhogdt} gave the optical
response for a system of two-level atoms.  For the case of
multilevel atoms the response follows from the general equations of
motion~\eqref{eq:cavityeqnmotionpsig} and~\eqref{eq:cavityeqnmotionpsie},
resulting in the coupled equations for the set of all possible
polarizations and coherences
\begin{subequations}
\begin{align}
{d\over dt}\,&\Pvhat^+_{\nu\eta} =
i\bar{\Delta}_{g\nu e\eta}\Pvhat^+_{\nu\eta}
\nonumber \\
&-\frac{{\cal D}^2}{\kappabar}{\sf P}^{\nu\eta}_{\eta\zeta}{\sf
G}_{c}(x,x)\dv_{g\zeta e\tau}\cpsi_{g\nu}\dpsi_{e\tau} \nonumber \\
&-i{\cal D}^2{\sf P}^{\nu\eta}_{\tau\nu}\gv(x)\cpsi_{e\tau}\dpsi_{e\eta}\hat{a}_F
+i{\cal D}^2{\sf P}^{\nu\eta}_{\eta\tau}\gv(x)\cpsi_{g\nu}\dpsi_{g\tau}\hat{a}_F
\nonumber \\
&-i{\cal D}^2{\sf P}^{\nu\eta}_{\tau\nu}\hv(x)\cpsi_{e\tau}\dpsi_{e\eta}
+i{\cal D}^2{\sf P}^{\nu\eta}_{\eta\tau}\hv(x)\cpsi_{g\nu}\dpsi_{g\tau}
\nonumber \\
&+\frac{{\cal D}^2}{\kappabar}\INT{x'} {\sf P}^{\nu\eta}_{\tau\nu}
{\sf G}_{c}(x,x')\,\cpsi_{e\tau}\Pvhat^+(x')\dpsi_{e\eta} \nonumber \\
&-\frac{{\cal D}^2}{\kappabar}\INT{x'} {\sf P}^{\nu\eta}_{\eta\tau}
{\sf G}_{c}(x,x')\,\cpsi_{g\nu}\Pvhat^+(x')\dpsi_{g\tau}\,, \label{eq:Pcavityfull} \displaybreak[0] \\
{d\over dt}\,&\cpsi_{g\nu}\dpsi_{g\eta} =
i\bar{\Delta}_{g\nu g\eta}
\cpsi_{g\nu}\dpsi_{g\eta}\nonumber \\
&+2\mbox{Re}\left[\frac{1}{\kappabar}\right]\dv_{e\zeta g\nu}\cdot{\sf G}_{c}(x,x)
\dv_{g\eta e\tau}\cpsi_{e\zeta}\dpsi_{e\tau} \nonumber \\
&-i\gv(x)\cdot\dv_{e\tau g\nu}\cpsi_{e\tau}\dpsi_{g\eta}\hat{a}_F
+i\gv^*(x)\cdot\dv_{g\eta e\tau}\hat{a}_F^\dagger\cpsi_{g\nu}\dpsi_{e\tau}\nonumber \\
&-i\hv(x)\cdot\dv_{e\tau g\nu}\cpsi_{e\tau}\dpsi_{g\eta}
+i\hv^*(x)\cdot\dv_{g\eta e\tau}\cpsi_{g\nu}\dpsi_{e\tau} \nonumber \\
&+\frac{1}{\kappabar}\INT{x'}\dv_{e\tau g\nu}\cdot{\sf G}_{c}(x,x')
\,\cpsi_{e\tau}\Pvhat^+\dpsi_{g\eta} \nonumber\\
&+\frac{1}{\kappabar^*}\INT{x'}\dv_{g\eta e\tau}\cdot{\sf G}^*_{c}(x,x')
\,\cpsi_{g\nu}\Pvhat^-\dpsi_{e\tau} \, , \label{eq:rhogcavityfull} \displaybreak[0] \\
{d\over dt}\,&\cpsi_{e\nu}\dpsi_{e\eta} =
i\bar{\Delta}_{e\nu e\eta}
\cpsi_{e\nu}\dpsi_{e\eta}
\nonumber \\
&-\frac{1}{\kappabar}\dv_{e\eta g\zeta}\cdot{\sf G}_{c}(x,x)\dv_{g\zeta
e\tau}\cpsi_{e\nu}\dpsi_{e\tau} \nonumber \\
&-\frac{1}{\kappabar^*}\dv_{g\zeta e\nu}\cdot{\sf G}^*_{c}(x,x)\dv_{e\tau
g\zeta}\cpsi_{e\tau}\dpsi_{e\eta} \nonumber \\
&+i\gv(x)\cdot\dv_{e\eta g\tau}\cpsi_{e\nu}\dpsi_{g\tau}\hat{a}_F
-i\gv^*(x)\cdot\dv_{g\tau e\nu}\hat{a}_F^\dagger\cpsi_{g\tau}\dpsi_{e\eta} \nonumber \\
&+i\hv(x)\cdot\dv_{e\eta g\tau}\cpsi_{e\nu}\dpsi_{g\tau}
-i\hv^*(x)\cdot\dv_{g\tau e\nu}\cpsi_{g\tau}\dpsi_{e\eta}\nonumber \\
&-\frac{1}{\kappabar}\INT{x'}\dv_{e\eta g\tau}\cdot{\sf
G}_{c}(x,x')\,\cpsi_{e\nu}\Pvhat^+(x')\dpsi_{g\tau}\nonumber \\
&-\frac{1}{\kappabar^*}\INT{x'}\dv_{g\tau e\nu}\cdot{\sf
G}^*_{c}(x,x')\,\cpsi_{g\tau}\Pvhat^-(x')\dpsi_{e\eta}\,,
\label{eq:rhoecavityfull}
\end{align}
\label{eq:cavityfullopticalresponse}
\end{subequations}
where the repeated indices $\zeta$ and $\tau$ should be implicitly summed over,
and $\bar{\Delta}_{a\nu b\eta} = \Delta_{b\eta}-\Delta_{a\nu}$.  Here we
explicitly indicate only the non-local position dependence of the atomic field
operators, and we have also introduced the convenient tensor
\beq
{\sf P}^{\nu\eta}_{\mu\tau} \equiv {\dv_{g\nu e\eta}
\dv_{e\mu g\tau}\over {\cal D}^2}
= \sum_{\sigma,\varsigma}
\pol_{\sigma}\pol_{\varsigma}^* {\cal C}_{\nu,\eta}^{(\sigma)}{\cal
C}_{\tau,\mu}^{(\varsigma)}
\,.
\label{pro}
\eeq
Taking expectation values of Eqs.~\eqref{eq:cavityfullopticalresponse}  gives
equations of motion for all one-body correlation functions, and knowledge of
those correlation functions solves the problem of optical response.  However,
as in the two-level case, the one-body correlation functions depend in turn
upon two-body correlation functions, leading to the hierarchy of equations of
motion for multi-level atoms.  The number of correlation functions involved in
the hierarchy increases rapidly with the number of levels present in the atoms,
and the complexity of the multi-level case can therefore be substantially
greater than in the two-level case.

\section{Hierarchy of equations for the optical response of atoms in a cavity}
\label{app:cavityhierarchy}

\subsection{Limit of low light intensity}

In the limit of low light intensity, Eqs.~\eqref{eq:cavityeqnmotionpsig}
and~\eqref{eq:cavityeqnmotionpsie}, along with commutators such as
\eqref{eq:cavitycommutator}, may be used to write the hierarchy of coupled
integral equations describing the optical response of an ensemble of two-level
atoms in an optical cavity in the compact form
\begin{widetext}
\begin{align}
  \dot{P}_\ell&(x_1,\ldots,x_{\ell-1}; x_\ell) = \left[i\bar{\Delta}  -
  \frac{\Dc^2}{\kappabar}G_c(x_\ell,x_\ell)\right]P_{\ell}
  (x_1,\ldots,x_{\ell-1}; x_\ell) +
  i\Dc^2\left[h(x_\ell)+a_F g(x_\ell)\right]
  \rhoF_\ell(x_1,\ldots,x_\ell) \nonumber\\
  &  -  \frac{\Dc^2}{\kappabar}\sum_{k=1}^{\ell-1} G_c(x_\ell,x_k)
  P_\ell(x_1,\ldots,x_{k-1},x_{\ell},x_{k+1},\ldots,x_{\ell-1};x_k)
   -  \frac{\Dc^2}{\kappabar}\INT{x_{\ell+1}} G_c(x_\ell,x_{\ell+1})
  P_{\ell+1}(x_1,\ldots,x_\ell;x_{\ell+1}) \, ,
  \label{eq:cavitylowintensityhierarchy}
\end{align}
analogous to the free-space case of Refs.~\cite{Ruostekoski1997a}
and~\cite{Lee16}.
Here we have neglected all terms of greater than first order in the light field
amplitude of excited state operators. Similar to
Ref.~\cite{Lee16}, we have defined the $\ell^{\mathrm{th}}$ order
one-dimensional correlation functions in the limit of low light intensity as
\begin{align}
\label{eq:CorrFuncDef_cavity}
P_\ell(x_1,\ldots,x_{\ell-1}; x_\ell) &\equiv
\quantmean{\hat\psi_g^\dag(x_1)\ldots\hat\psi_g^\dag(x_{\ell-1})\hat{P}^+(x_\ell)
  \hat\psi_g(x_{\ell-1})\ldots\hat\psi_g(x_1)}\,,\\
\rhoF_\ell(x_1,\ldots,x_\ell) &\equiv
\quantmean{\hat\psi_g^\dag(x_1)\ldots\hat\psi_g^\dag(x_{\ell})
  \hat\psi_g(x_{\ell})\ldots\hat\psi_g(x_1)} \,,
\label{eq:DensityCorrFuncDef_cavity}
\end{align}
\end{widetext}
and again used the definitions $\hat{P}^+ = \Dc\cpsi_g\dpsi_e$, $h(x)
=\hv(x)\cdot\dv_{eg}/{\cal D}$, $g(x) =
\gv(x)\cdot\dv_{eg}/{\cal D}$, and $G_c(x,x') = g(x)g^*(x')$.

Equation~\eqref{eq:cavitylowintensityhierarchy} shows that, as in the free space
case, the $\ell^{\mathrm{th}}$ order correlation function $P_\ell$ depends on
the integral over $P_{\ell+1}$, leading to a hierarchy of equations of motion
which terminates only at $\ell=N$.  Avoiding the need to solve this hierarchy of
equations, the stochastic technique presented in
Sec.~\ref{sec:cavitytwolevel} gives a computationally efficient
method to obtain the optical response.  For the low light intensity case
corresponding to Eq.~\eqref{eq:cavitylowintensityhierarchy}, the stochastic
method solves the linear set of equations~\eqref{eq:cavitydiscretelowlight} for
stochastic realizations of fixed atom positions. An argument akin to that in
App. B of Ref.~\cite{Lee16} shows that subsequent averaging over an ensemble
of such realizations reproduces the full dynamics of the correlation functions
as dictated by Eq.~\eqref{eq:cavitylowintensityhierarchy}.

\subsection{Including saturation for a system of two atoms}

Going beyond the limit of low light intensity to include the effects of saturation
rapidly increases the complexity of the hierarchy of equations of motion governing
the correlation functions, since one must now account for all $4^n$ $n$-body
correlation functions for every $n < N$.  However, for a small system of just two
atoms the hierarchy terminates with the $16$ equations of motion for the two-body
correlation functions $\av{\cpsi_i(x)\cpsi_j(x')\dpsi_k(x')\dpsi_m(x)}$, where
$i,j,k,m$ can refer to the ground or excited atomic state.  In such a system the
direct steady-state solution of the full hierarchy for the optical response is
therefore a realistic, if somewhat tedious, approach.

The equations of motion for the two-body correlation functions may be derived
using Eqs.~ (\ref{eq:cavityeqnmotionpsig}) and (\ref{eq:cavityeqnmotionpsie}),
and terms reordered using commutators similar to
Eq.~(\ref{eq:cavitycommutator}).  For example, the equations of motion for the
correlation functions $P_2(x;x') =
\Dc\av{\cpsi_g(x)\cpsi_g(x')\dpsi_e(x')\dpsi_g(x)}$  and $\av{\cpsi_g(x)\cpsi_e(x')\dpsi_g(x')\dpsi_e(x)}$
are governed by
\begin{widetext}
\begin{align}
\frac{d}{dt}&\av{\cpsi_g(x)\cpsi_g(x')\dpsi_e(x')\dpsi_g(x)} =
\left(i\bar{\Delta}(x') -\frac{\Dc^2}{\kappabar}G_c(x',x')\right)\av{\cpsi_g(x)\cpsi_g(x')\dpsi_e(x')\dpsi_g(x)}
\nonumber \\
&+ \Dc^2G_c^*(x,x')\left(\frac{1}{\kappabar}+\frac{1}{\kappabar^*}\right)
\av{\cpsi_g(x)\cpsi_e(x')\dpsi_e(x')\dpsi_e(x)}
+ \Dc^2G_c(x,x)\left(\frac{1}{\kappabar}+\frac{1}{\kappabar^*}\right)
\av{\cpsi_e(x)\cpsi_g(x')\dpsi_e(x')\dpsi_e(x)}
\nonumber \\
&-\frac{\Dc^2}{\kappabar}G_c(x',x)\av{\cpsi_g(x)\cpsi_g(x')\dpsi_g(x')\dpsi_e(x)}
\nonumber \\
&+i\Dc\left[a_F^* g^*(x)+h^*(x)\right]\av{\cpsi_g(x)\cpsi_g(x')\dpsi_e(x')\dpsi_e(x)}
-i\Dc\left[a_Fg(x')+h(x')\right]\av{\cpsi_g(x)\cpsi_e(x')\dpsi_e(x')\dpsi_g(x)}
\nonumber \\
&-i\Dc\left[a_F g(x)+h(x)\right]\av{\cpsi_e(x)\cpsi_g(x')\dpsi_e(x')\dpsi_g(x)}
+i\Dc\left[a_F
g(x')+h(x')\right]\av{\cpsi_g(x)\cpsi_g(x')\dpsi_g(x')\dpsi_g(x)}\,,
\label{eq:eqnmotionP2cavity2atoms} 
\end{align}
\begin{align}
\frac{1}{\Dc^2}\frac{d}{dt}&\av{:\Pop^+(x)\Pop^-(x'):}
=\frac{d}{dt}\av{\cpsi_g(x)\cpsi_e(x')\dpsi_g(x')\dpsi_e(x)}\\
=&
\left\{i\left[\bar{\Delta}(x)-\bar{\Delta}(x')\right]-{\Dc^2}\left(\frac{G_c(x,x)}{\kappabar}+\frac{G_c(x',x')}{\kappabar^*}\right)\right\}
\av{\cpsi_g(x)\cpsi_e(x')\dpsi_g(x')\dpsi_e(x)}
\nonumber \\
&+\Dc^2G_c(x,x')\left(\frac{1}{\kappabar}+\frac{1}{\kappabar^*}\right)
\av{\cpsi_e(x)\cpsi_e(x')\dpsi_e(x')\dpsi_e(x)}
\nonumber \\
&-\frac{\Dc^2}{\kappabar^*}G_c(x,x')
\av{\cpsi_e(x)\cpsi_g(x')\dpsi_g(x')\dpsi_e(x)}
-\frac{\Dc^2}{\kappabar}G_c(x,x')\av{\cpsi_g(x)\cpsi_e(x')\dpsi_e(x')\dpsi_g(x)}
\nonumber \\
&+i\Dc\left[a_F^*g^*(x')+h^*(x')\right]\av{\cpsi_g(x)\cpsi_e(x')\dpsi_e(x')\dpsi_e(x)}
-i\Dc\left[a_F^*g^*(x')+h^*(x')\right]\av{\cpsi_g(x)\cpsi_g(x')\dpsi_g(x')\dpsi_e(x)}
\nonumber \\
&-i\Dc\left[a_Fg(x)+h(x)\right]\av{\cpsi_e(x)\cpsi_e(x')\dpsi_g(x')\dpsi_e(x)}
+i\Dc\left[a_Fg(x)+h(x)\right]\av{\cpsi_g(x)\cpsi_e(x')\dpsi_g(x')\dpsi_g(x)}
\,.\label{eq:eqnmotionP+P-2atoms}
\end{align}
Deriving all $16$ such equations of motion, solving for the steady state, and
including the atom number conservation relation
\begin{align}
\rhoG_2(x,x') &= \av{\cpsi_g(x)\cpsi_g(x')\dpsi_g(x')\dpsi_g(x)} +\av{\cpsi_g(x)\cpsi_e(x')\dpsi_e(x')\dpsi_g(x)}\nonumber \\
&+\av{\cpsi_e(x)\cpsi_g(x')\dpsi_g(x')\dpsi_e(x)}+\av{\cpsi_e(x)\cpsi_e(x')\dpsi_e(x')\dpsi_e(x)},
\label{eq:cavityhierarchy_atomnumbercons}
\end{align}
\end{widetext}
leads to a set of $16$ linear independent equations which can be solved
numerically to give the set of all two-body correlation functions.  Here
$\rhoG_2(x,x')$ is the two-body density-density correlation function that we
assume is known. Consistent with our earlier stochastic treatment, we have
assumed that the atoms are stationary, consequently $\rhoG_2(x,x')$ is invariant
and has a form dictated by the trapping potential, and atom statistics. Once the two-body correlation functions are obtained, polarization
and excited state densities then follow from the steady state of the expectation
values of Eqs.~(\ref{eq:cavitydPdt}) and~(\ref{eq:cavitydrhogdt}), and the
hierarchy of equations has then been solved.

\end{document}